\documentclass{aa}
\usepackage[varg]{txfonts}
\usepackage{natbib}
\usepackage{amsmath}
\usepackage{multirow}
\usepackage{graphicx}
\usepackage[normalem]{ulem} 
\bibpunct{(}{)}{;}{a}{}{,}

\DeclareMathOperator*{\argmin}{arg\,min}

\begin{document}

\title{Machine learning techniques to select \textit{Be} star candidates} 
\subtitle{An application in the OGLE-IV Gaia south ecliptic pole field}

\author{M. F. P\'erez-Ortiz\inst{1}\fnmsep\inst{2}\fnmsep\inst{3}, A. Garc\'ia-Varela\inst{1}, 
A. J. Quiroz\inst{2}, B. E. Sabogal\inst{1}, and J. Hern\'andez\inst{4}}
\institute{Universidad de los Andes, Departamento de F\'{\i}sica,
  Cra. 1 No. 18A-10, Bloque Ip, A.A. 4976, Bogot\'a, Colombia \\
  \email{josegarc@uniandes.edu.co, 
bsabogal@uniandes.edu.co, aj.quiroz1079@uniandes.edu.co}
\and 
Universidad de los Andes, Departamento de Matem\'aticas,
Cra. 1 No. 18A-10, Edificio H, Bogot\'a, Colombia
\and Korteweg-de Vries Institute for Mathematics, University of Amsterdam, Science Park 105-107, 1098 XG Amsterdam,\\ The Netherlands 
\and Instituto de Astronom\'ia, Universidad Nacional Aut\'onoma de M\'exico, Unidad Acad\'emica en Ensenada, \\Ensenada BC 22860, M\'exico.
}  

\date{Received May 12, 2016; accepted June 23, 2017}

\abstract {Optical and infrared variability surveys produce a large
     number of high quality light curves.  Statistical pattern
     recognition methods have provided competitive solutions
     for variable star classification at a relatively low
     computational cost. In order to perform supervised
     classification, a set of features is proposed
     and used to train an automatic classification
     system. Quantities related to the
     magnitude density of the light curves and their Fourier
     coefficients have been chosen as features in
       previous studies. However, some of these features are not
       robust to the presence of outliers and the
     calculation of Fourier coefficients is computationally expensive
     for large data sets.} 
{We propose and evaluate the performance of a new robust set
       of features using  supervised classifiers in order to look for 
       new \textit{Be} star candidates in the OGLE-IV Gaia
       south ecliptic pole field.}
{We calculated the proposed set of features on six types of
     variable stars and also on a set of \textit{Be}
     star candidates reported in the literature.  We evaluated
     the performance of these features using classification trees and 
     random forests along with the K-nearest neighbours, support vector
     machines, and gradient boosted trees methods. We tuned the classifiers with a
       10-fold cross-validation and grid search. We then validated
       the performance of the best classifier on a
     set of OGLE-IV light curves and applied this to find new
     \textit{Be} star candidates.}
{The random forest classifier outperformed the others. By using the random forest classifier 
and colours criteria we found 50 \textit{Be} 
star candidates in the direction of the Gaia south
     ecliptic pole field,  four of which have infrared colours that are consistent 
     with Herbig Ae/Be stars.}
{Supervised methods are very useful in order to obtain preliminary samples
of variable stars extracted from large databases. As usual, the stars classified  
as \textit{Be} stars candidates must be checked for the colours and   
spectroscopic characteristics expected for them.}

\authorrunning{M. F. P\'erez-Ortiz et al.}  
  
\maketitle

\section{Introduction}
In the last 30 years, several photometric surveys have
been releasing huge amounts of data. This has motivated the
use of statistical and computational techniques to process and analyse
these large data sets (\citet{Bass}, \citet{Pichara},
and references therein) and generate many 
catalogues of variable stars. 
\textit{Be} stars are a particular class of variables, which  despite more than 100 years of their discovery, evolutionary state, and dependency on metallicity are 
yet under study \citep{Ri}.  For this reason, samples 
of 
\textit{Be} stars  in different environments are needed, 
and consequently,  methods to classify the stars in a systematic way as well. \\\\

\citet{debosscher_automated_2007} and \citet{sarro_automated_2009}
proposed using supervised learning methods to classify light
curves of variable stars. This approach is a three-step process:
representation, training, and evaluation. Light curves are represented
with a set of features. These features can be categorical,
discrete, or continuous parameters that are calculated for each light
curve. They have to be informative enough to identify with high
probability the variability class to which each light curve belongs. 
In the training step, a learning algorithm is used to infer, from
available previously classified data (a training sample), a rule that
assigns to each point of the feature space a variability type, 
that is, a classifier.  Then, this rule can be used to classify
light curves that have not been previously used in the
training step. Finally, in the evaluation step, the
performance of the resulting classifier is assessed on data
that were not used in the training stage.

\begin{figure*}[!ht]
  \centering
  \includegraphics[scale=1]{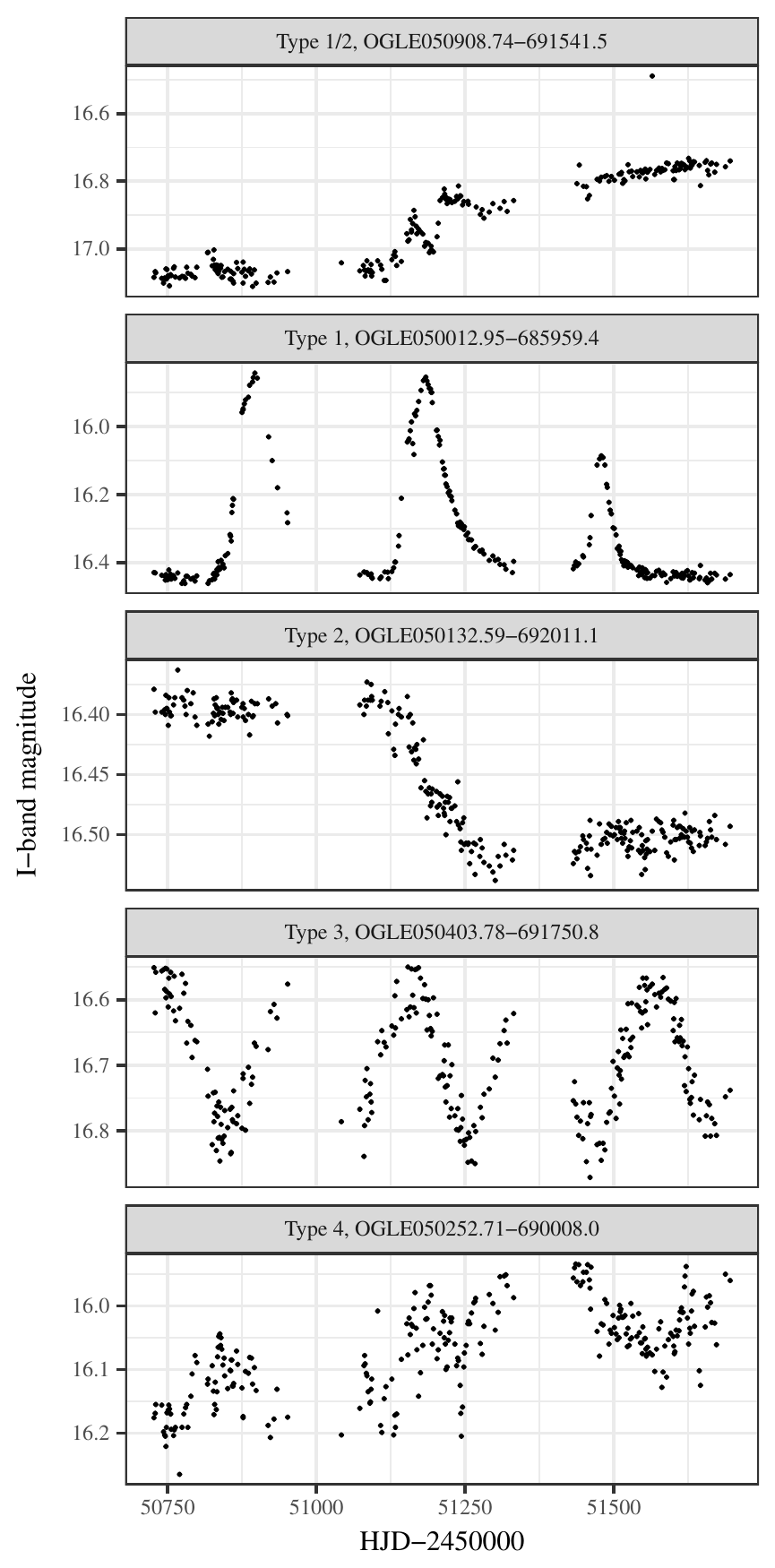}
  \caption{OGLE-II  time series of LMC BeSC. Observations
    were sampled in a window close to 900 days.}
  \label{figure:graf2}
\end{figure*}

The selection of features is crucial because it is the only 
information available to the classifier. 
\citet{debosscher_automated_2007} proposed using
Fourier coefficients of light curves as features, finding that they
could be used to classify classical Cepheids, Mira, RR
Lyr\ae, among other variable stars. \citet{deb_light_2009} performed principal
component analysis on the interpolated values of magnitudes
  after folding the light curves using their periods. These
authors found that the dimensionality of the representation of the
light curves could be greatly
reduced. \citet{park_classification_2013} used the multi-scale 
visualisation technique, called thick-pen
transformation, on the folded light curve to obtain features that can
be used for classification. \citet{kim_epoch_2014} used a set of
features that included the period of the light curves, quantities
derived from Fourier decomposition, descriptive
statistics of the magnitude density, and colour indexes. Despite the
existence of efficient algorithms performing Fourier analysis,
computing the Fourier coefficients is a demanding task for 
large data sets and the periods computed by automatic procedures
often need to be checked manually.
\\\\
\textit{Be} stars are non-super giant very rapid rotators with
spectral types between late O and early A, whose spectra at some time show or have
shown one or more Balmer lines in emission, which are generated in a circumstellar
decretion disk that emerges from the ejection of stellar mass whose
causes are yet under study (\citealt{CO}, \citealt{Ri}). The decretion disk is conceptually 
different from the accretion disk that can be observed around young 
stellar objects such as Herbig Ae/Be (HAeBe) stars. The accretion disks 
are optically thick  and  feed a central young star. \\
\textit{Be} stars show irregular spectroscopic
and photometric variability. This behaviour is called the \textit{Be}
phenomenon. The most complete description of
\textit{Be} stars until now, including observations and models, has
been presented by \citet{Ri}. Photometric searches for \textit{Be}
star candidates (BeSC from hereafter) 
and the subsequent spectroscopic follow-up are useful to obtain 
samples of \textit{Be} stars that allow us to analyse and prove
different scenarios of the \textit{Be} phenomenon. In
particular, \citet{MEN2} performed a photometric search for BeSC
within the Small Magellanic Cloud (SMC) with the OGLE-II variable star catalogue.  Those authors found that light curves of BeSC have
morphologies similar to those of classical \textit{Be} stars, but also
they found other BeSC with completely different morphologies
with diverse light curves.  Based on the long-term
morphology, those authors reported five types of variability: Type-1
stars are objects showing outbursts, some of which are characterised by a
rise of brightness followed by a gradual decline lasting tens
of days; and others are characterised by more symmetric rising and fading
timescales, lasting hundreds of days. Their amplitudes are
  about 0.2 mag. The Galactic \textit{Be} stars $\lambda$ Eri, $\mu$
Cen, and those stars reported by \citet{HU8} and \citet{HU}, exhibit
this kind of variability.  Type-2 stars show a brightness
discontinuity or jump of the order of a few tenths of magnitudes 
that occurs on timescales of about few hundreds of days.
This behaviour had never been observed in Galactic \textit{Be} stars,
as was also confirmed by \citet{sabogal_search_2014}. 
Type-3 stars show periodic or quasi-periodic magnitude variations.
Type-4 stars are objects with light curves showing stochastic
magnitude variations, such as those exhibited by classical
\textit{Be} stars. \citet{MEN2} also mentioned a group of BeSC light
curves that showed brightness jumps and outbursts
simultaneously. These stars were classified as Type1/2. Figure \ref{figure:graf2} shows examples of these 
morphological types.  \citet{Sabogal2005} also found these morphological behaviour 
of BeSC in the Large Magellanic Cloud (LMC). 
Despite the diversity of
 the shape of their light curves, these stars are collectively
  classified as BeSC. Other light curve examples can be found in \citet{MEN2} and
  \citet{sabogal_search_2014}.  

Following the ideas of
\citet{sabogal_search_2014}, who used descriptive statistics of the
magnitude density to search for BeSC, in this work we 
propose and evaluate a new set of features to classify variable 
stars and particularly BeSC.
Since light curves sometimes 
contain atypical measurements, our descriptive statistics 
needs to be robust to the presence
  of such measurements.\ We train
a set of state-of-the-art classifiers on a subset of six types of
variable stars selected from OGLE-III and a set of BeSC to verify the usefulness of the
  features for performing classification of variables. 
Subsequently, we validate
the resulting classifiers on a data set from the OGLE-IV database
and use the best performing classifier to look for BeSC.

This article is organised as follows. In section \ref{sect:data}, we
describe the data used to train the classifiers. In section
\ref{sect:features} we describe our features, the random forests classifier and
  classification trees.  We only report the
  classification results for these two methods.  
  For the sake of brevity of the main text, we
  defer the description of the other automatic classification
  techniques that we consider to the Appendix. In section
\ref{sect:results}, we show our results and discuss our findings 
of classification of variable stars using random forests and
  classification trees.  In section \ref{sect:besearch} 
we present the results of applying the random forests 
  method to classify BeSC from the OGLE-IV Gaia south
ecliptic pole field catalogue (hereafter \textit{OGC}). 
In section 6 we present a brief study of the infrared colours of
our BeSC. Finally, in section \ref{sect:conclusions} 
we present the main conclusions of this work.

\section{Data\label{sect:data}}

This work makes use of the variable star catalogues of the  
OGLE project, a long-term experiment whose main
objective is searching for dark matter via gravitational lensing. This project began in 1992 and is in its fourth phase since 2010.  The
observations of this project have been made with the 1.3 m Warsaw
telescope at Las Campanas Observatory in
Chile. Characteristics of the new 32 chip mosaic camera and a
  technical overview can be found in \citet{Udalski2015}.
  
Two OGLE data sets are used in this work.\ The first comes
  from OGLE-III and \citet{sabogal_catalogue_2008}. We use this data set to
  train and test our classifiers. The
  second comes from OGLE-IV. We use this data set to validate the best
  performing classifier. Within this last data set we also search for BeSC.

The OGLE-III data set consists of 432333 I band light curves 
\citep{udalski_optical_2004} of variable stars\footnote{\url{ftp://ftp.astrouw.edu.pl/ogle/ogle3/OIII-CVS}}  
belonging to the GB and Magellanic Clouds. 
These data cover about eight years from 2001 to 2009 \citep{Udalski2015}.
From these data we select Cepheids
(Ceph), $\delta$ Scuti ($\delta$ Sct), eclipsing binaries (EB), long
period variables (LPV), RR Lyr\ae{} (RR Lyr), and type II Cepheid
(T2Ceph) as our training sample. These variability classes 
and the number of stars in each
class are shown in Table \ref{table:data}. Additionally, we 
selected the OGLE-III I band light curves of 475
BeSC reported by \citet{sabogal_catalogue_2008} in the direction of
the GB, and of 200 BeSC reported by \citet{Sabogal2005} in the
LMC. These 675 BeSC are included in our training 
sample since they clearly exhibit the five morphological 
types shown  in Figure \ref{figure:graf2}.

\begin{table}[ht]
  \centering  
  \caption{Training data set.}
      \label{table1}
  \begin{tabular}{llllc}
    \hline
    \hline
    Variability & \multirow{2}{*}{Location} & Number & \multirow{2}{*}{Total} & \multirow{2}{*}{References} \\
    type        &     & of objects & &   \\ 
    \hline
    BeSC        & GB  & 475    &       & 1 \\
                & LMC & 200    &  675  & 18\\
    \hline     
                & GB  & 32     &       & 2 \\
    Ceph        & LMC & 3344   & 8006  & 3 \\
                & SMC & 4630   &       & 4 \\
    \hline
    $\delta$ Sct& LMC & 2788   & 2788  & 5 \\
    \hline
    \multirow{2}{*}{EB} 
                & LMC & 26121  & \multirow{2}{*}{32259} &  6 \\
                & SMC & 6138   &       & 7 \\
    \hline
                & GB  & 232406 &       & 8 \\
    LPV         & LMC & 91995  & 343785& 9 \\
                & SMC & 19384  &       & 10\\
    \hline
                & GB  & 16836  &       & 11\\
    RR Lyr      & LMC & 24906  & 44217 & 12\\
                & SMC & 2475   &       & 13\\
    \hline
                & GB  & 357    &       & 14, 15 \\
    T2 Ceph     & LMC & 203    &   603 & 16\\ 
                & SMC & 43     &       & 17\\
    \hline
  \end{tabular}
  \tablebib{
    (1) \citet{sabogal_catalogue_2008};  
    (2) \citet{soszynski_optical_2011};
    (3) \citet{soszynski_optical_2008-1};
    (4) \citet{soszynski_optical_2010-2};
    (5) \citet{poleski_optical_2010};
    (6) \citet{graczyk_optical_2011};
    (7) \citet{pawlak_eclipsing_2013-1};
    (8) \citet{soszynski_optical_2013-1};
    (9) \citet{soszynski_optical_2009};
    (10) \citet{soszynski_optical_2011-1};
    (11) \citet{soszynski_optical_2011-2};
    (12) \citet{soszynski_optical_2009-1};
    (13) \citet{soszynski_optical_2010};
    (14) \citet{soszynski_optical_2011}; 
    (15) \citet{soszynski_optical_2013};
    (16) \citet{soszynski_optical_2008};
    (17) \citet{soszynski_optical_2010-1};   
    (18) \citet{Sabogal2005}
    }
  \label{table:data}  
\end{table}

We applied the best performing classifier on the second data
set, which consists of 6789 I band light curves. These light curves were
reported and catalogued by \citet{soszynski_optical_2012} in the study of
OGLE-IV variable stars\footnote{\url{ftp://ftp.astrouw.edu.pl/ogle/ogle4/GSEP/var_stars}}
in the Gaia south ecliptic pole (GSEP) field. 
This work reported Ceph, $\delta$ Sct, EB, LPV, RR
Lyr, and T2Ceph, as shown in Table \ref{table:dataOGLEIV}. Stars
showing variability with characteristics different to the mentioned
classes, or showing similar characteristics with ambiguities, were
assigned by \citet{soszynski_optical_2012} to the class
"Other", within this class 19 BeSC are reported. However,
a visual inspection of the light curves belonging to the
Other class suggests that the number of BeSC could be larger. For this
reason we decide to look for BeSC in this data set.

\begin{table}[h]
  \centering  
  \caption{Gaia south ecliptic pole field OGLE-IV variables}
    \label{table:dataOGLEIV}
    \resizebox{88mm}{!}{ 
  \begin{tabular}{cccccccc}
    \hline
    \hline
    Type &  Ceph  & $\delta$ Sct & EB & LPV & RR Lyr & T2 Ceph & Other \\
    \hline    
    Number & 135 & 159 & 1532 & 2799 & 686 & 5 & 1473\\
    \hline
  \end{tabular}
   }   
\end{table}

\section{Set of features \label{sect:features}}
In this section we describe what we
  mean by robustness; then we discuss our approach toward
  calculating robust quantities, describe our set of features, and
  visualise the data in the resulting feature space. \\
 
  As OGLE variability studies are made principally in the I band, 
we compute the features only in that band. Our set of features carry information about the I band time
series and are robust to the presence of outlying values, that is,
their values do not change dramatically in the presence of such
measurements as opposed to their non-robust counterparts. The
robustness of a statistic is usually measured with the so-called
breakdown point. This is the fraction of the data that needs to be
contaminated before the statistic takes arbitrarily high (or low)
values \citep[see][Chap. 1]{huber_robust_2009}. We use the word
``robust'' in that sense throughout the document.

The approach we choose to calculate robust quantities
  is not the only approach in existence. One might be inclined to use a
two-step process of first finding outlying values in the magnitude
series and then applying classical estimates of parameters instead of
their robust counterparts. We prefer the use of robust estimators for
the following reasons. First, the process of automatic outlier
identification in complex data is prone to false rejections and false
retentions. For instance, popular outlying detection techniques could
identify high magnitude values in the light curve of an EB
system as outliers when they are not. Second, the process of screening
for outliers and then applying classical statistical estimators to the
remaining data usually requires the employment of robust estimators
for the outlier identification step. Finally, robust estimation
methods deal with outliers by appropriately down-weighting their
effect on the resulting estimators. For a more detailed discussion,
see \citet{hampel_robust_1986} or \citet{staudte_robust_1990}.

In their study, \citet{sabogal_search_2014} used kurtosis and skewness. Kurtosis is a measure of both peakedness and tail weight. Skewness is the third standardised moment. 
In this work, we use the measure of skewness proposed by \citet{brys_robust_2004},
  the octile skewness (OS) along with the measures of tail weight proposed
  by \citet{brys_robust_2006}, the left octile weight (LOW), and the
  right octile weight (ROW). We do not use the kurtosis and skewness
  because their calculation involve the third and fourth power of the
  deviation of the data points from the mean, which makes them very
  sensitive to outlying values. On the other hand the robustness of
OS, LOW, and ROW comes from the robustness of quantile estimators. The OS
is defined by
\begin{equation}
  OS = \frac{(Q_{0.875}-Q_{0.5})-(Q_{0.5}-Q_{0.125})}{Q_{0.875}-Q_{0.125}},
\end{equation}
where $Q_{p}$ is the \textit{p} quantile of the magnitude distribution, that
is, the value of $I$, such that the fraction $p$ of the values of I is
smaller than $Q_{p}$. The OS is the difference between the lengths of the
right and the left tails of the distribution scaled so that its
maximum value is 1. 
It is positive for right-skewed distributions and
negative for left-skewed distributions. Similarly,
\begin{equation}                                                        
  LOW = \frac{(Q_{0.375}-Q_{0.25})-(Q_{0.25}-Q_{0.125})}{Q_{0.375}-Q_{0.125}}, 
\end{equation}
and
\begin{equation}                                                          
  ROW = \frac{(Q_{0.875}-Q_{0.75})-(Q_{0.75}-Q_{0.625})}{Q_{0.875}-Q_{0.625}}, 
\end{equation}
describe how heavy the tail (left or right) of the distribution is
relative to its magnitude near the centre of the distribution.  

As estimators of location and scale we choose the
median and the median absolute deviation (MAD), respectively. The
median is $Q_{0.5}$ and MAD is defined by
\begin{equation}
  MAD = \mathrm{median}_i(|I_i-\mathrm{median}_j(I_j)|),
\end{equation}
where $I_i$ is the $i$-th value of magnitude of the light
  curve in question. The $MAD$ is a measure of the dispersion of the
magnitude distribution.

\begin{figure}[!ht]
  \centering
    \includegraphics[scale = 1]{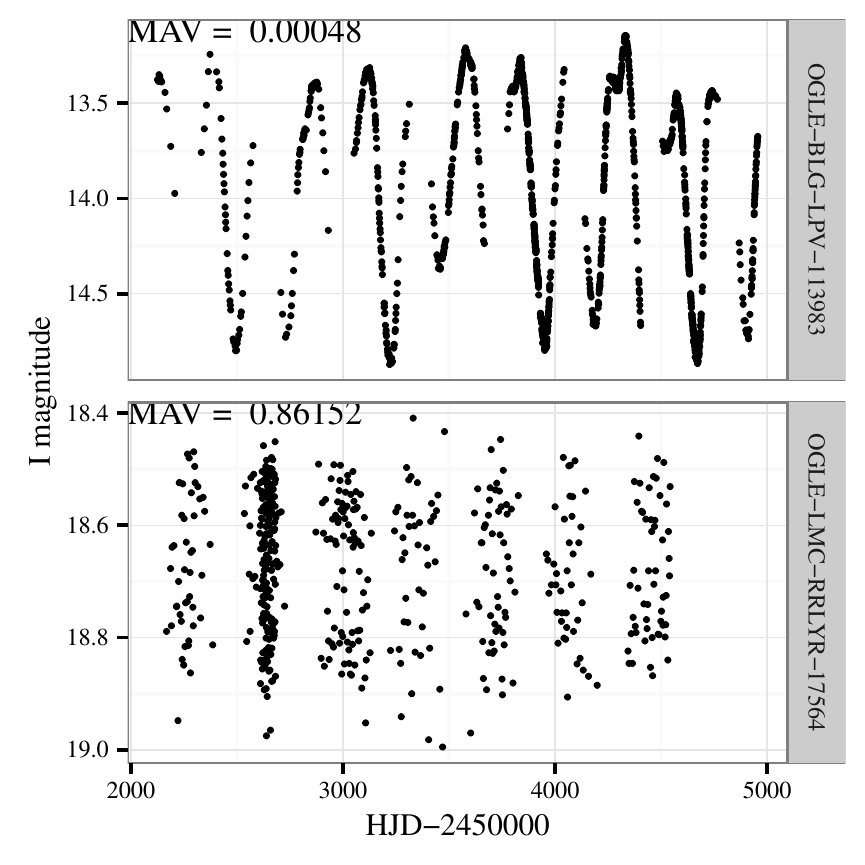}
    \caption{Time series of two variable stars from the OGLE-III
      catalogue. Upper panel shows LPV data with low MAV value,
      corresponding to curves that vary in the same timescale as the
      measurements. Bottom panel shows RR Lyr data with a higher MAV value
      that looks random before being folded.}
  \label{figure:abbeDifferences}
\end{figure}

To measure the smoothness of the light curves, we choose a modified
version of the Abbe value ($\mathcal{A}$)  originally
proposed by \citet{von_neumann_distribution_1941} and later used by
\citet{mowlavi_searching_2014} in the search of transients. The Abbe value is
defined by
\begin{equation}
  \label{eq:abbeValue}
  \mathcal{A} = 
  \frac{n}{2(n-1)}\frac{\sum_{i=1}^{n-1}(I_{i+1}-I_{i})^2}{\sum_{i=1}^n(I_{i}
    -\overline{I})^2}
\end{equation}
and compares the quadratic increments $(I_{i+1}-I_{i})^2$ with the
standard deviation of the light curve. The $\mathcal{A}$ tends to one for
a purely noisy light curve and to zero when the light curve shows a
high degree of smoothness. In the case of periodic curves, when the
sampling frequency is small with respect to the frequency of the
curve, the light curve looks random before being folded and the
Abbe value is close to one. This means that in this case the Abbe
value does not reflect the smoothness of the folded curve, which is a
limitation.  On the other hand, for those curves whose variation
patterns can be seen using the unfolded light curve, $\mathcal{A}$
is small. Since the quantities, $\overline{I}$, and
$(I_{i}-\overline{I})^2$ are sensible to the presence of outlying
values, we choose to modify this value by repeatedly using a robust
measure of location instead of averages. As robust measure of location
we use the M estimator proposed by \citet{huber_robust_1964} and explained in
\citet{venables_modern_2013}. For a set of points $y_1,\dots,y_N$,
Huber's estimate of location is the point $x$ at which
\begin{equation}
  \label{eq:huberLocEstimate}
  \sum_{i=1}^{N}\rho\left(\frac{y_i-x}{MAD}\right)
\end{equation}
is minimised for
\begin{equation}
  \label{eq:huberFunction}
  \rho(x) = \begin{cases} 
    -c &\mbox{if } x < -c \\ 
    x & \mbox{if } |x|<c\\
    c & \mbox{if } x>c, 
  \end{cases}
\end{equation}
where $MAD$ is the median absolute deviation of the $y_i$ and
  $c$ can be chosen freely. We use $c = 1.5$, that is, we winsorise
at $1.5$ of the MAD. The point $\mathrm{loc}_i(y_i)$ at which equation 
(\ref{eq:huberLocEstimate}) is minimised is often called 
a Winsorised mean, defined by
\begin{equation}
  \label{eq:loc}
  \mathrm{loc}_i(y_i) = 
\argmin_{x}\sum_{i=1}^{N}\rho\left(\frac{y_i-x}{MAD}\right)
\end{equation}
in the statistical literature.
We then propose as measure of smoothness the modified Abbe value (MAV) 
\begin{equation}
  \label{eq:modifiedAbbe}
  MAV = \frac{1}{2} \: \frac{\mathrm{loc}_i((I_{i+1}-I_{i})^2)}{\mathrm{loc}_{i}((I_i-\mathrm{loc}_jI_j)^2)},
\end{equation}
which has properties similar to those of $\mathcal{A}$. In figure
\ref{figure:abbeDifferences}, we compare two light curves with different values of
$MAV$.

The proposed feature statistics, i.e. the median, OS, LOW, ROW, MAD,
  and MAV combine robustness with the ability to measure the skewness,
  tail weight, location, scale, and smoothness of the light curves.
In the case of OS, LOW, and ROW, their breakdown
value is $12.5\%$, while that of the MAD and the median is $50\%$,
which is the highest possible. In the case of OS, LOW, and ROW, we
find that this level of robustness is enough for our purposes since
it is rare to find light curves with such high levels of
contamination. Other quantities with a higher breakdown point
would result in less sensibility to distributional changes in tail
weight and skewness \citep[]{brys_robust_2004, brys_robust_2006}.

\begin{table}[ht]
  \centering
  \caption{Statistical features}
  \begin{tabular}{lr}
    \hline
    \hline
    Measurement & Robust quantity  \\
    \hline
    Location & Median \\ 
    Scale &  Median absolute deviation (MAD)\\
    Skewness & Octile skewness (OS) \\
    Tail weight & Left octile weight (LOW) \\
    & Right octile weight (ROW)  \\
    Smoothness & Modified Abbe value (MAV)\\
    \hline
  \end{tabular}
  \label{table:usedFeatures}
\end{table}

\begin{figure}
  \centering
    \includegraphics[scale = 1]{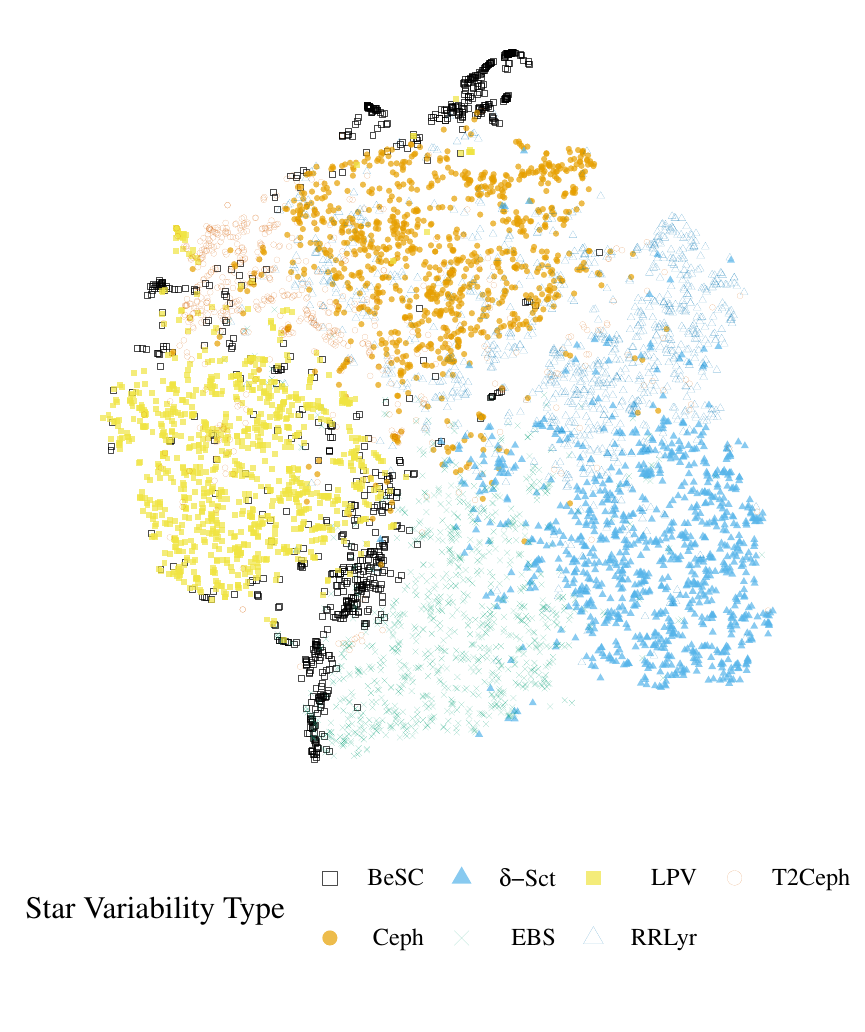}
    \caption{Visualisation of the six-dimensional feature space by
      t-SNE technique. The axes are omitted because the scale 
        and orientation of this embedding carries no meaning. It can
      be seen how the different variability classes are separated in
      the six-dimensional feature space, although some overlapping of
      the classes is also evident.}
  \label{figure:tsne}
\end{figure}

\begin{figure}
  \centering
    \includegraphics[scale = 1]{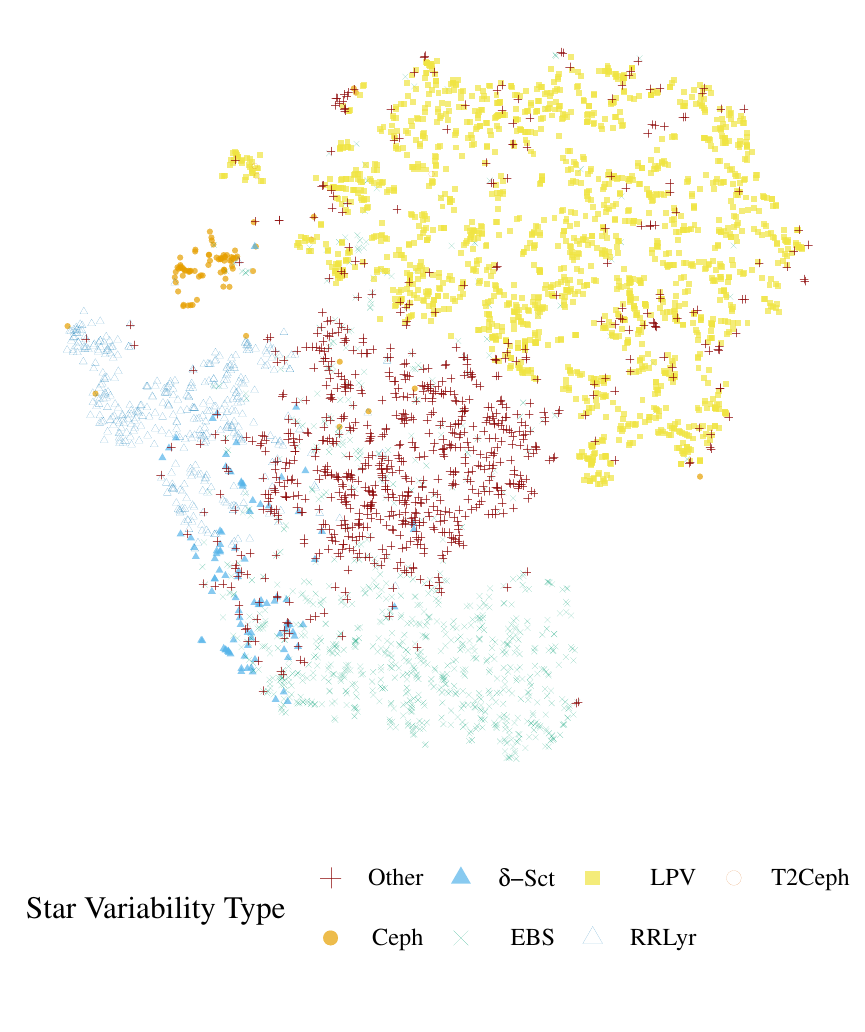}
    \caption{t-SNE visualisation of the OGC data. The axes are omitted
      because the scale and orientation of this embedding
      carries no meaning.}
  \label{figure:tsneOGLEIV}
\end{figure}

In order to visualise how the data look in in the resulting
  six-dimensional feature space, we use the t-distributed stochastic
neighbour embedding (t-SNE) \citep[]{van_der_maaten_visualizing_2008}.
This is a non-supervised visualisation technique (it does not use the
class to which each point belongs) that seeks to embed the six-dimensional data in the plane. This is carried out by minimising a loss
function that captures the discrepancy between the high-dimensional and the two-dimensional structure. This
procedure only involves one free parameter, called perplexity, which
is to be set by the user. The perplexity is a continuous measure of
how many neighbours of each point are taken into account. Figure
\ref{figure:tsne} shows a t-SNE plot of the data used for
  training and Figure \ref{figure:tsneOGLEIV} shows a t-SNE plot of
  the data from OGLE-IV GSEP field. We built these figures  via the
\texttt{Rtsne} package \citep[][]{rtsne_package} for R and a
perplexity value of $40$. The data set used for Figure \ref{figure:tsne}
is a random sample that consists of either all or 2000 points of each
variability class. This sample was subsampled further to avoid
overplotting. In this plot it can be readily seen that light
  curves of the same variability class cluster together and that there
  is overlap between light curves from different classes that have
  similar light curves, for example T2Ceph and Ceph. The structure of this
plot is robust to the choice of random sample.

\subsection{Method for classifier performance evaluation
    and parameter selection} \label{sect:classificationError}

Each classifier considered has one or more parameters that
need to be tuned to maximise their performance. For each
classifier, we performed a grid search on the classifier parameter space,
that is, for each point in a grid of parameters, we estimated
  the performance and then chose one classifier. In this section we
describe the procedure that we used to estimate the performance of the classifiers.  We
used normalised confusion matrices to prevent the imbalance in the number of light
curves that belong to each class from affecting the performance measures. For more information on model
selection via repeated cross-validation,
\citet{hastie_elements_2009} can be consulted.

We use 10-fold cross-validation and normalised confusion matrices to
estimate the performance of the models that we consider. We randomly divided the data into 10 parts, called folds, with roughly the same
number of examples, using stratified random sampling so that each
fold had the same proportion of light curves of each variability
class. For each fold, standard cross-validation was
performed. This means that in each iteration the data on the
fold were used as a hold-out sample on which a classifier trained with
the remaining data was tested. In each iteration we cross-tabulated
the observed and predicted classes for all light curves in the
fold. This cross-tabulation is called (un-normalised)
confusion matrix. We call the confusion matrix for the k-th iteration
$C^{(k)}$. This is a table where each column represents the instances of
the class to which each light curve belongs, while each row represents
the class to which the classifier assigned it. This means that the
entry $C_{ij}^{(k)}$ of the confusion matrix $C^{(k)}$ is the number
of elements of the class $j$ that were classified as belonging to the
class $i$ in the k-th   cross-validation iteration. Thus, the cases in
the diagonal are those that are classified correctly, while the
cases outside the diagonals show the number of times that each
possible error occurred. We normalised the confusion matrix so that
each column sums up to one for reasons that we now explain.

Since the data that we used are surely not representative of the star
populations that are observed, the use of the un-normalised confusion
matrix can lead to misleading results. For instance, one common
performance metric is the accuracy, which gives the proportion of
light curves that are correctly classified. This can be the wrong
measure of performance because of the imbalance in the number of
members in each class.  For example, if a classifier decided that all
light curves of our sample are LPV, it would have almost 80\% accuracy
in our sample simply because of the over-representation of that
variability class. Such a classifier achieves a relatively high
accuracy but is not of practical use. Two other popular measures of
performance are recall (also called sensitivity) and precision (also
called positive predictive value). Recall is the proportion of light
curves that are correctly classified, given that the light curves
belong to a particular variability class.  For the l-th class and the
k-th cross-validation iteration, the recall is
\begin{equation}
  \label{eq:recall} 
  \textrm{recall}_l^{(k)} = \frac{C_{ll}^{(k)}}{\sum_{i}C_{il}^{(k)}},
\end{equation}
and precision is the fraction of light curves that are correctly
classified given that the light curves are classified as belonging to
a particular variability class. For the l-th class and the k-th cross-validation iteration, the precision is
\begin{equation}
  \label{eq:precision}
  \textrm{precision}_l^{(k)} = \frac{C_{ll}^{(k)}}{\sum_{j}C_{lj}^{(k)}}.
\end{equation}
The cross-validation estimate of recall is 
\begin{equation}
  \label{eq:cvRecall}
  \textrm{recall}_l=\frac{1}{10}\sum_{k}\textrm{recall}_{l}^{(k)}
\end{equation}
and is an estimator of the conditional probability of
correct classification given that a light curve belongs to a
variability class, while the cross-validation estimate of precision is
\begin{equation}
  \label{eq:cvPrecision}
  \textrm{precision}_l=\frac{1}{10}\sum_{k}\textrm{precision}_{l}^{(k)}
\end{equation}
and is an estimator of the posterior probability of a
light curve belonging to a particular class given that it is
classified as such. While recall is not affected by the
unrepresentativeness of our sample, precision is thus affected. For instance, if
just 1\% of LPV light curves were classified by a hypothetical model
as BeSC and no other element of other class is wrongly classified as
such, BeSC precision would drop to 12\% when all BeSC light curves are
correctly classified just because of the large number of LPV light
curves in our sample. To avoid this and because the real proportion of
objects belonging to each class in the observed fields is unknown, we
normalised the confusion matrices by setting the population of each
class to one, so that each column of the confusion matrices sums up to
one. This is, if we call $\widehat{C}^{(k)}$ the normalised
  confusion matrix in the k-th fold, then
  \begin{equation}
    \widehat{C}^{(k)}_{ij} = \frac{C_{ij}^{(k)}}{\sum_{i}C_{ij}^{(k)}}.
  \end{equation}
  We estimated precision, and recall analogues of equations
  (\ref{eq:recall}) to (\ref{eq:cvPrecision}) using
  normalised confusion matrices. This can be shown to lead to
  consistent estimates of the class conditional and posterior
  probability of correct classification (see appendix
  \ref{appendix:UniformPrior}), when the a priori
  probabilities of the different variability classes are all equal,
  that is, when the a priori distribution of the classes is uniform.

Ideally, a model should achieve perfect recall and precision for each
class, but in practice it is found that there is a trade off between
these two quantities when tuning parameter models: a compromise should
be achieved. For this purpose, we use the mean $F_1$
score. For each class, the F1 score is defined as the harmonic
  mean of the precision and the recall 
\begin{equation}
  F_{1,l} =
  \frac{1}{ \frac{1}{\textrm{precision}_{l} }+\frac{1}{
      \textrm{recall}_{l}} }
  = 2 \: \frac{\textrm{precision}_{l}\times\textrm{recall}_{l}}{ 
\textrm{precision}_{l}+\textrm{recall}_{l}}.
\end{equation}
The corresponding estimator from the k-th cross-validation
  iteration is
\begin{equation}
  F_{1,l}^{(k)} = 
2 \: \frac{\textrm{precision}_{l}^{(k)}\times\textrm{recall}_{l}^{(k)}}{\textrm{
precision}_{l}^{(k)}+\textrm{recall}_{l}^{(k)}}.
\end{equation}
The mean (over the variability classes) F1 score for the k-th
  fold is
\begin{equation}
  \overline{F}_{1}^{(k)} = \frac{1}{7}\sum_{l}F_{1,l}^{(k)}
\end{equation}
and, finally, the cross-validation estimator of the mean F1
  score is
\begin{equation}
  \overline{F}_{1} = \frac{1}{10}\sum_{k}\overline{F}_{1}^{(k)}.
\end{equation}

In the case of tree-based algorithms, in which the parameters
of the classifier control their complexity, we choose the simplest
model whose cross-validation estimate of the mean $F_1$ score is
within one standard deviation (over the folds) of the highest value
\citep[Chap. 7]{hastie_elements_2009}. This is carried out because simpler
models are preferred and, in this case, the performance of the best
and the simplest model cannot be statistically distinguished.


\subsection{Classification trees and random forests}\label{sect:classifiers}

In this subsection we discuss in certain detail the random
  forest (RF) classifier, which achieved the best performance in our
  evaluation. We also describe classification trees (CT), which is
  necessary to understand random forests. Other classifiers considered
  in our study are described in Appendix \ref{appendix:otherclassifiers}. 
  All of the classifiers considered
  are non-linear, state-of-the-art classifiers, which are described, for
instance, in \citet{hastie_elements_2009}. We use implementations of
the classifiers in the R-statistical computing environment
\citep{rcore2015}, and the wrapper and other useful functions provided
in the classification and regression training (caret) package
\citep[][]{wing_caret:_2016}. Parameters for each classifier are tuned
up following the process described in the previous subsection.

\subsubsection{Classification trees  \label{sect:MethClassificationTrees}}

Classification trees were first proposed by Breiman, Friedman, Stone, and
Olshen throughout several works that were later summarised by
\citet{breiman_classification_1984}. The decision rule is implemented
in the form of a binary decision tree. At each node a simple question
is asked about one feature, and at the terminal nodes, a class is
assigned to each example. In the case of numerical features $x_i$,
these questions are of the form $x_i\leq c$ for constants $c$ that are
chosen during the training step. These trees are constructed from the
root by successively dividing the data using binary questions that
maximise the reduction of a measure of ``impurity'', that is, the
diversity of classes in the resulting nodes. This process of
successive division is repeated until each node contains only a
predefined minimum number of examples or until they are pure. The
resulting tree is usually large and, in order to avoid
  over fitting the data, it is then trimmed to reduce its complexity
and improve its general properties. The resulting tree can be
interpreted easily, since the divisions in the feature space give
insight to the characteristics of the light curves.

We used the implementation of CT provided in the
\texttt{rpart} package \citep[][]{therneau_rpart:_2015}. In this
implementation, a complexity parameter (\texttt{cp})  needs to be
tuned. It is the minimum decrease in the re-substitution estimate
error that each partition has to achieve. The re-substitution
  estimate of the tree error is obtained from the proportions of the
  data classes at the terminal nodes and the a priori probability of
  each class \citep[][]{breiman_classification_1984}, which we chose
  to be uniform.

\subsubsection{Random forests}

Random forests were proposed by \citet{breiman_random_2001} based on the
idea that a set of weak classifiers can vote to conform a strong
robust classifier. This method consists of building a large number of
classification trees whose decisions are not very correlated and then
taking the majority vote among them as the decision of the random
forest. Each classification tree is built with a random sample taken
with replacement from the complete learning sample (bootstrap
sampling) using a random subset of a fixed size of features to reduce
the correlation among trees.  Each of the trees considered may over fit
the data, but the ensemble does not, so pruning becomes
unnecessary. Nevertheless, smaller trees may be grown by limiting
their size. Thus, only the number of features that are randomly chosen
for each tree, the total number of trees, and their size need to be
tuned. \citet{biau_consistency_2008} showed that the decision rule
given by random forests converges to the best possible decision rule
for a given set of features when the size of the training set
$N\to\infty$.

We used the implementation of RF provided in the
\texttt{randomForest} package \citep[][]{liaw_classification_2002}.
We tuned the number of trees (\texttt{ntree}), the maximum number of
nodes of each tree (\texttt{max\_nodes}), and the number of features
randomly chosen for each tree (\texttt{mtry}).

\section{Results and discussion \label{sect:results}}

For CT and RF, the two classifiers performing best, we report
the normalised confusion matrix, which we estimate using 10-fold
cross-validation. We also report the cross-validation estimates of the
recall and precision for each class with their cross-validation
standard deviation (CV SD), that is, their standard deviation over the
folds. In Table \ref{table:f1Scores}, we report the estimates of the
mean $F_1$ scores for the five classifiers considered. We
find the RF classifier to achieve the best performance.

In general, the decisions of the classifiers are not related in a
simple manner. For classifiers with similar performances,
  their decisions are usually correlated. This happens because in
some regions of feature space, where there are predominantly objects
of one class, most well-performing classifiers agree on their
decision, while in other regions, where there exists a mixed
proportion of objects belonging to different classes, the performance
of a classifier depends heavily on the complexity and shape of
the decision boundary that they can learn. At the same time, the
shape of the decision boundaries depends on the sample size and parameter
choice. When comparing RF and CT, we find that they agreed on
93\% of the sample. The majority of these stars are LPV objects, where
both classifiers perform well. In the case of T2Ceph objects, where CT
perform better than RF, 84\% of the objects correctly classified by CT
are also correctly classified by RF, and in the rest of classes, the
agreement is higher than 94\%.

Now we give a description of tuning and preprocessing steps that we
follow CT and RF, as well as a brief discussion of their
  performance. We also present the results of applying the RF
classifier to the Other sample in the OGC.

\begin{table}[ht]
  \centering
  \caption{Mean $F_1$ scores of the tuned classifiers.}
  \begin{tabular}{lcc}
    \hline
    \hline
    Classifier & Mean $F_1$ Score & CV SD \\
    \hline
    Random forests           & 0.86 & 0.01 \\
    Classification trees     & 0.81 & 0.01 \\
    Gradient boosted trees   & 0.75 & 0.02 \\
    Support vector machines  & 0.72 & 0.02 \\
    K-nearest neighbours     & 0.65 & 0.01 \\
    \hline
  \end{tabular}
  \label{table:f1Scores}
\end{table}

\subsection{Classification trees}

We assessed the performance of CT with uniform prior on four values of the complexity parameter, i.e. $cp = 0.1, 0.01, 0.001, 0.0001$
(see section \ref{sect:MethClassificationTrees}).  We finally set the
complexity parameter to 0.001 because lowering this parameter further than this
does not bring statistically significant improvements and drastically increases the complexity of the resulting trees. In Table
\ref{table:resultsClassificationTrees} we show the performance of the
resulting classifier. The CT classifier offers a good compromise in terms of
  recall and precision, when compared to other classifiers, with the
  exception of RF. Ceph, RR Lyr, and T2 Ceph objects are classified
  with low recalls and sensitivities, but they are mainly confused
  between each other because the light curves of these objects
  are very similar.

\begin{table}[ht]
\centering
\caption{Cross-validation results of classification trees.}
\resizebox{88mm}{!}{
  \begin{tabular}{lccccccc}
     & \multicolumn{7}{c}{Reference} \\
    \hline\hline
    Prediction   & BeSC & Ceph & $\delta$Sct & EBS & LPV & RRLyr & T2Ceph \\ 
    \hline
    BeSC         & 0.93 & 0.02 &      & 0.02 & 0.02 &      & 0.01 \\ 
    Ceph         & 0.01 & 0.77 &      &      &      & 0.18 & 0.13 \\ 
    $\delta$Sct  &      & 0.01 & 0.89 & 0.06 &      & 0.03 &      \\ 
    EBS          & 0.01 & 0.02 & 0.05 & 0.86 & 0.01 & 0.03 & 0.04 \\ 
    LPV          & 0.04 & 0.01 &      & 0.02 & 0.94 &      & 0.06 \\ 
    RRLyr        &      & 0.05 & 0.04 & 0.01 &      & 0.69 & 0.06 \\ 
    T2Ceph       & 0.01 & 0.13 & 0.00 & 0.02 & 0.03 & 0.06 & 0.70 \\ 
    \hline
    Recall       & 0.93 & 0.77 & 0.89 & 0.86 & 0.94 & 0.69 & 0.70 \\ 
    CV SD        & 0.03 & 0.02 & 0.02 & 0.01 & 0.01 & 0.01 & 0.08 \\
    \hline
    Precision    & 0.93 & 0.70 & 0.90 & 0.84 & 0.87 & 0.81 & 0.74 \\ 
    CV SD        & 0.02 & 0.03 & 0.01 & 0.03 & 0.03 & 0.03 & 0.02 \\ 
   \hline
    Number       & 675  & 8006 & 2788 & 32259&343785& 44217& 603 \\
    \hline 
    \label{table:resultsClassificationTrees}
  \end{tabular}
}
\end{table}

\subsection{Random forests \label{section:resultsRF}}

We used a uniform prior and assess the performance
of this method on five values of the number of features randomly selected
for each tree, $mtry = 2,3,4,5$ and grew a forest with 100, 200, and
500 trees without pruning. We find that different values of $mtry$
do not affect the performance of the method and that
growing more than 200 trees does not have a significant effect
on our performance metrics. Results for unpruned trees were
not satisfactory, so we modified the maximum number of
terminal nodes \texttt{max\_nodes} that each tree in the forest could
have. Since in the previous experiments the values of \texttt{mtry}
did not affect the performance of the model, we fixed
\texttt{mtry} to 2, and tried 10 values of
\texttt{max\_nodes}: $2^1, 2^2, \dots, 2^{10}$ while \texttt{ntree}
was held fixed at 200. The maximum mean $F_1$ score
was achieved at \texttt{max\_nodes}$=2^9$ and
\texttt{max\_nodes}$=2^{10}$. We set the maximum number of nodes to
$2^9$ and obtained the results shown in Table
\ref{table:resultsRandomForestMN9}. This model achieved a
better overall performance with recall/precision of 0.92/0.97 for BeSC
objects; 0.91/0.91 for $\delta$-Scuti objects; 0.99/0.86 for LPV
objects. Sensitivities for RRLyr, T2Ceph, and Ceph are 
lower than in the case of CT, but these classes were
again confused among them. Additionally, since the
maximum number of nodes \texttt{max\_nodes} is smaller than that of
unpruned trees, computation time and the memory needed is reduced.

\begin{table}[ht]
\caption{Cross-validation results of random forest. 
Each tree has a maximum number of nodes equal to $2^9$.}
\centering
\resizebox{88mm}{!}{
  \begin{tabular}{lrrrrrrr}
    & \multicolumn{7}{c}{Reference} \\
    \hline\hline
    Prediction  & BeSC & Ceph & $\delta$Sct & EBS & LPV & RRLyr & T2Ceph \\
    \hline
    BeSC        & 0.92 &      &      & 0.01 &      &      &  \\ 
    Ceph        & 0.01 & 0.91 & 0.01 &      &      & 0.20 & 0.16 \\ 
    $\delta$Sct &      &      & 0.91 & 0.05 &      & 0.04 &  \\ 
    EBS         & 0.02 & 0.01 & 0.05 & 0.89 &      & 0.02 & 0.03 \\ 
    LPV         & 0.04 & 0.01 &      & 0.02 & 0.99 &      & 0.09 \\ 
    RRLyr       &      & 0.02 & 0.03 & 0.01 &      & 0.72 & 0.06 \\ 
    T2Ceph      & 0.01 & 0.04 & 0.00 & 0.01 &      & 0.03 & 0.66 \\  
    \hline
    Recall      & 0.92 & 0.91 & 0.91 & 0.89 & 0.99 & 0.72 & 0.66 \\ 
    CV SD       & 0.04 & 0.01 & 0.02 & 0.01 & $\sim 10^{-3}$ & 0.01 &
    0.06 \\
    \hline
    Precision   & 0.97 & 0.71 & 0.91 & 0.87 & 0.86 & 0.85 & 0.88 \\ 
    CV SD       & 0.01 & 0.02 & 0.01 & 0.02 & 0.03 & 0.03 & 0.03 \\ 
\hline
    Number       & 675  & 8006 & 2788 & 32259&343785& 44217& 603 \\
    \hline

    \label{table:resultsRandomForestMN9}
  \end{tabular}
}
\end{table}

\subsection{Validation on the OGLE-IV Gaia south ecliptic pole field
  data \label{subsect:OGLEIVvalidation}}

Random forest is the classifier that achieved the best overall
performance during the cross-validation process in the OGLE-III
data. We trained a RF classifier using  the complete
OGLE-III data set and the optimal
parameters found in Section \ref{section:resultsRF}. We tested the resulting classifier on the
OGC obtaining the results shown in Table
\ref{table:resultsOGLEIV}. Since the variability classes of the new
test data do not coincide with those of the training data, only recall
(the proportion of objects that belong to a specific class that are
correctly classified) for the classes found in both data sets is
reported. No information about the classification posterior
distribution can be extracted. The lowest recall is achieved for EB
and T2Ceph objects at 72\% and 60\%, respectively. In the sample there
are only five T2Ceph objects and four are classified either as T2Ceph or
Ceph, but because of the small number of examples of this class, this
result needs to be interpreted with caution. Remarkably, the rest of
the objects belonging to the rest of variability classes, Ceph,
$\delta$Sct, LPV, and RRLyr, are correctly classified at rates higher
than 90\%.  From the Other class, 108 objects are classified as BeSC,
and of those, 19 are the objects that were identified previously as
BeSC by \citet{soszynski_optical_2012}. Besides, our classifiers
found that in the Other class there are EB, $\delta$Sct, 
and LPV as shown in Table \ref{table:resultsOGLEIV}.

Despite the differences between the OGC and OGLE-III data in time
span ($\sim$8 yr versus $\sim$ 2.4 yr) and the average number of 
photometric points per stars in the I band ($\sim$100-3000 versus $\sim$ 340), 
these results suggest that our set of features can also be used in such
situations. Also, the proportion of time series belonging to each
class in the OGC is not similar to that of the OGLE-III data either,
where the imbalance from an abundance of LPV objects is much
larger, or to the uniform prior distribution used to train the
RF classifier. These results suggest that the procedure of
assigning a uniform prior distribution and our set of features may
be well suited for this situation and that the over-representation of
LPV objects can be effectively overcome with these procedures.
Nevertheless, since we do not have any other sensible prior
distribution to which to compare, further conclusions could not be
reached.

The RF classifier assigns to each object the class most
  frequently selected by the trees that makes up the forest. Table
  \ref{table:majorityOctiles} shows the size of that "majority" in our
  RF classification of the OGC data, by giving the octiles of the
  percentages of trees choosing the assigned class. For instance, half
  of the time, the assigned class gets the vote of at least $96\%$ of
  the trees, while $87.5\%$ of the classifications are made with a
  majority of at least $57\%$ of the trees. In general, the assigned
  class is selected by an ample majority of trees, especially
  considering that the votes are split among seven different classes, in
  principle.  The Other class, as reported by
  \citet{soszynski_optical_2012}, contains objects whose variability
  type could not be unambiguously determined. This class includes objects
  that resemble rotating spotted stars, BeSC, and other
  variables. Since the training stage of the RF classifier did   not include objects with the characteristics of some of those stars, RF probably assigns an incorrect class to some of those stars.  
  This is a shortcoming of applying the supervised learning  methods.

\begin{table}[ht]
\caption{Results of the random forest classifier on the
OGLE-IV data set.}
  \centering
  \resizebox{88mm}{!}{
    \begin{tabular}{lrrrrrrr}
      & \multicolumn{7}{c}{Reference} \\
      \hline\hline
      Prediction & Ceph & $\delta$Sct & EB & LPV & RRLyr & T2Ceph & Other\\ 
      \hline
      BeSC        & 1   &     & 21   & 3    &     &   & 108 \\ 
      Ceph        & 126 & 1   & 42   &      & 10  & 1 & 52  \\ 
      $\delta$Sct &     & 146 & 209  &      & 9   &   & 316 \\ 
      EB          &     & 2   & 1110 & 2    & 13  &   & 676 \\ 
      LPV         & 1   &     & 105  & 2790 &     & 1 & 226 \\ 
      RRLyr       & 3   & 10  & 19   &      & 652 &   & 86  \\ 
      T2Ceph      & 4   &     & 26   & 4    & 2   & 3 & 9   \\ 
      \hline
      Total       & 135 & 159 & 1532 & 2799 & 686 & 5 & 1473\\ 
      \hline
      Recall      & 0.93& 0.92& 0.72 & 0.99 & 0.96& 0.60 & - \\ 
    \label{table:resultsOGLEIV}
    \end{tabular}
  }
\end{table}

\begin{table}[ht] 
  \centering
  \caption{Octiles of the size of the majority in RF classification. }
   \resizebox{88mm}{!}{ 
  \begin{tabular}{ccccccccc}
    \hline\hline
    {Octile} & 100 & 87.5 & 75 & 62.5  & 50 & 37.5 & 25 & 12.5\\
    Majority Size (\%)  & 24.5  & 57.0 & 77.5  & 90.5  & 96.0 & 98.5 & 99.5 & $\sim$100\\
    \hline
  \end{tabular}}
  \label{table:majorityOctiles}
\end{table}

\section{Looking for BeSC in the OGLE-IV Gaia south ecliptic pole
  field using random forests \label{sect:besearch}}

A visual inspection of the light curves classified as Others
by \citet{soszynski_optical_2012} suggests the presence of additional BeSC
in the Other sample than the 19 reported by the authors. Since
  the RF classifier achieved the highest $F_1$ score for BeSC objects
  among the five classifiers considered, we used to look for BeSC in this data
  set.  Additionally, we trained a binary RF classifier that
distinguished BeSC objects from non-BeSC objects using the
OGLE-III data and an a priori probability of 0.5. In order to
  train this binary RF, we followed the same procedure that we
  used for training the multi-class RF classifier, but we do not report
  the results because of the following.  By inspecting the
  position of the objects that were classified as BeSC by both
  classifiers in the colour-magnitude diagram compared to the
rest of the OGC sample, we decided to select the multi-class
RF for further analysis because we believe it to be less prone to
produce false positives. The multi-class RF classifier selects
  108 objects as BeSC, while the binary RF classifier selects 215
  objects. The multi-class RF classifier recovers the 19 BeSC
reported previously in the OGC, and both the multi-class and
  the binary classifiers coincide in selecting 100 objects as
  BeSC, of which 18 had been previously reported in the OGC.

\begin{figure}[!ht]
  \centering
    \includegraphics[width=\hsize]{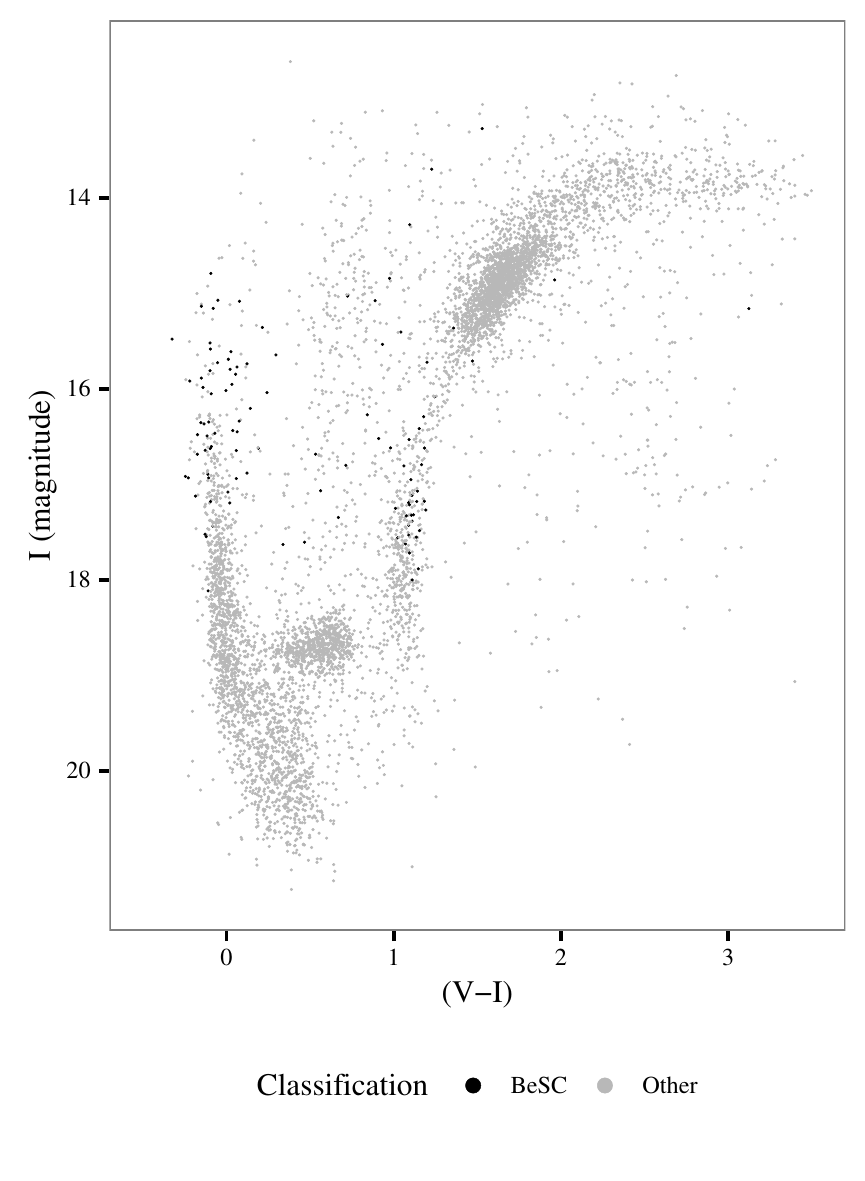}
  \caption{OGC colour-magnitude diagram for about 6700 variable stars 
  reported by  \citet{soszynski_optical_2012}. Black points  
  represent 108 stars  selected as BeSC by our multi-class RF classifier.
Median magnitudes are used instead of average magnitudes.}
  \label{figure:cmd}
\end{figure}

Figure \ref{figure:cmd} shows the colour-magnitude diagram of all
variable stars with $(V-I)$ colours reported by \citet{soszynski_optical_2012}.
The BeSC selected in this work
using the multi-class RF classifier are highlighted as darker
points on the diagram. Two distinct groups of stars classified as BeSC
by our procedure can be identified in Figure \ref{figure:cmd}. One of
these groups have stars showing blue colours, as expected to \textit{Be} stars.
The other is located in the red giant branch,
indicating that these stars could  probably be slowly pulsating
  variables (SPV) or LPV, whose light curve morphologies are similar
to these of \textit{Be} stars but their colours are redder.  In order to obtain
a more reliable list of BeSC, we discard the stars with colours out of the expected range of colours
for \textit{Be} stars  from those initially labelled
as BeSC.  Intrinsic colours for Galactic \textit{Be} stars,
including their typical infrared excess, has been reported to be
$-0.35 < (V-I)_{0} < 0.8$ mag \citep{Wisniewski}.  The GSEP field
covers four OGLE-IV fields \citep{soszynski_optical_2012}, three of
which are located about 270 arc min from the centre of the LMC. We search
for the colour excess values of these fields at the Galactic
Dust Reddening and Extinction Archive\footnote{On NASA/IPAC
    Infrared Science page, which uses the extinction maps and values
    reported by \citet{Schlafly}:
  http://irsa.ipac.caltech.edu/applications/DUST}.  For three fields
(LMC562, LMC563, and LMC570), a $E(V-I)$ value of 0.093 mag is
reported. These values are not derived from the IRAS/COBE extinction
maps,  while a value of 0.068 mag, for the field LMC571,  
was obtained from these maps. We adopt 0.093 mag as the colour 
excess for all fields.
Therefore, we select BeSC the stars within the colour
range $-0.257 < (V-I) < 0.893$ mag as more reliable, obtaining a total of 50
stars.
\begin{table}
 
\caption{Catalogue of BeSC in the GSEP field. Single or double asterisk 
appended to the last column indicates the stars that had been classified in the
OGC as BeSC or the stars with infrared colours consistent
with those of the HAeBe stars, respectively.}
\resizebox{88mm}{!}{
\centering
  \begin{tabular}{lccccl}
      \hline
       \hline
    ID   & RA & Dec &  V & (V-I) &  Type \\
    \hline
LMC562.19.8354 & 05:58:54.52 & -67:12:00.4 &   16.916   & -0.195  & Type-1\\
LMC563.21.7054 & 05:54:11.13 & -65:58:00.5 &   16.103     & -0.156  & Type-1*\\
LMC562.28.8855 & 05:56:36.44 & -66:56:51.4 &      16.660     & -0.134  & Type-1\\
LMC562.11.9588 & 05:57:04.69 & -67:31:03.2 &   16.357     & -0.126  & Type-1*\\
LMC562.14.124 &  05:52:19.95 & -67:39:59.4 &     16.286     & -0.122  & Type-2\\
LMC562.14.10726 &  05:51:55.37 & -67:31:07.9 &      16.874    & -0.119  & Type-1\\
LMC562.01.211 &  06:00:43.40 & -67:58:14.4 &  16.608     & -0.112  & Type-1*\\
LMC562.24.11360 &   05:51:09.41 & -67:12:14.0 &  16.483     & -0.107  & Type-1*\\
LMC562.13.11357  &  05:53:14.85 & -67:34:10.2 &     15.551    & -0.104  & Type-1\\
LMC562.16.231 &  05:48:48.80 & -67:43:37.8 &  16.918     & -0.097  & Type-1* \\
LMC562.27.92 &   05:59:07.72 & -67:05:13.6 &  16.612    & -0.091  & Type-1*\\
LMC562.26.8110 & 06:00:52.74 & -66:58:17.4 &  16.399    & -0.089 &   Type-1*\\
LMC562.02.7937 & 05:58:49.19 & -67:55:28.9 &     15.159     & -0.085  & Type-4\\
LMC562.06.10895 &  05:51:53.39 & -67:48:51.0 &    18.126     & -0.070 &   Type-1\\
LMC563.04.477 &  05:53:21.94 & -66:47:34.0 &     17.478     & -0.049 &   Type-4\\
LMC562.09.110 &  06:00:10.48 & -67:38:31.9 &    16.535     & -0.031 &   Type-1\\
LMC562.32.173 &  05:50:07.46 & -67:05:56.6 &     16.407     & -0.016 &   Type-4\\
LMC562.24.132 &  05:50:04.86 & -67:22:33.3 &  15.721     & -0.008 &  Type-1*\\
LMC562.13.11454 &  05:53:22.34 & -67:29:13.2 &  15.983    & 0.006 &   Type-1/2*\\
LMC562.21.180 &  05:55:28.96 & -67:26:44.3 &  16.894    & 0.025 & Type-1*\\
LMC563.16.113 &  05:48:08.49 & -66:31:02.6 &    16.883    & 0.042  & Type-3\\
LMC562.13.106 &  05:53:17.60 & -67:43:38.0 &    15.848  & 0.053   & Type-3\\
LMC563.06.110 &  05:50:01.51 & -66:46:23.4 &  16.340   & 0.063  & Type-1/2*\\
LMC562.13.11442 &  05:53:24.01 & -67:30:36.1 &    16.451   & 0.066  & Type-2\\
LMC562.12.10123  & 05:55:32.53 & -67:32:20.9 &    15.088    & 0.075 &   Type-4\\
LMC562.20.85 &   05:57:07.95 & -67:25:42.9 &    15.749    & 0.088  & Type-3\\
LMC562.15.132 &  05:50:21.21 & -67:41:43.8 &    15.717   & 0.111  & Type-1\\
LMC562.20.9119 & 05:57:20.00 & -67:12:46.5 &  16.065  & 0.113  & Type-1*\\
LMC563.30.7056 & 05:52:13.54 & -65:41:11.4 &    16.455    & 0.116  & Type-4\\
LMC562.01.7994 & 06:00:12.35 & -67:47:04.3 &    15.558  & 0.133  & Type-4\\
LMC562.03.8441 & 05:56:48.78 & -67:49:57.0 &    15.518   & 0.150  & Type-1\\
LMC562.24.11487 &  05:51:01.93 & -67:15:24.0 &    16.631   & 0.187  & Type-2\\
LMC562.04.125 &  05:56:02.78 & -67:57:41.8 &    16.122   & 0.191  & Type-1\\
LMC562.28.207 &  05:56:30.97 & -67:04:51.3 &    17.171   & 0.197  & Type-4\\
LMC562.15.11956 &  05:50:54.39 & -67:29:25.9 &  15.190    & 0.282  & Type-1*\\
LMC570.14.103 &  06:02:19.40 & -67:01:48.9  &  15.634   & 0.304 & Type-4\\
LMC571.20.3879 & 06:05:28.35 & -65:20:06.9 &    17.623   & 0.366  & Type-3\\
LMC562.11.87 &   05:57:19.89 & -67:43:19.9 &    15.695   & 0.370  & Type-2\\
LMC562.20.78 &   05:56:45.52 & -67:26:52.2 &  15.216  & 0.451  & Type-1*\\
LMC562.02.8135 & 05:59:12.24 & -67:50:07.5 &    17.503   & 0.484  & Type-2\\
LMC562.16.12173 &   05:48:19.89 & -67:29:40.6 &  15.112  & 0.580  & Type-1*\\
LMC563.04.129 &  05:53:57.47 & -66:50:01.6 &    17.078    & 0.582  & Type-2**\\
LMC563.17.142 &  05:59:42.70 & -66:09:08.0 &    16.543    & 0.610  & Type-1**\\
LMC562.07.11068 &  05:50:26.35 & -67:51:52.0 &    15.056   & 0.697  & Type-4\\
LMC562.25.11162 &  05:49:04.56 & -67:14:09.4 &    15.029   & 0.723 &   Type-4\\
LMC570.26.70 &   06:10:52.24 & -66:30:11.4 &    16.730    & 0.757  & Type-4**\\
LMC562.27.90 &   05:59:08.13 & -67:05:20.0  &  16.269   & 0.846  & Type-3\\
LMC570.17.266 &  06:12:12.82 & -66:46:06.0 &    17.330   & 0.849  & Type-2**\\
LMC563.05.450 &  05:52:48.46 & -66:47:56.5 &    16.519    & 0.877  & Type-1\\
LMC562.13.103 &  05:53:44.69 & -67:43:54.5 &  14.911   & 0.888  & Type-1*\\
    \hline
    \label{table:catalog}
  \end{tabular}
}
\end{table}

Table \ref{table:catalog} presents the catalogue of these BeSC. The first
column gives the OGC ID and the second and third columns show the equatorial
coordinates (J2000). The fourth column gives the I band magnitude of each
star. The fifth column shows the (V-I) colour for each star (all of these
data are taken from \citet{soszynski_optical_2012}). The last column
gives our classification of the light curves based on the morphological 
types described by \citet{MEN2}. The total number of stars of each of these
types is shown in Table \ref{table:num}.

\begin{table}[ht]
\caption{Types of BeSC found in the OGLE-IV GSEP field.}
  \begin{tabular}{ccccc}
       \hline
       \hline
    Type-1   & Type-2 & Type-3 &  Type-4 & Type-1/2  \\
    \hline
25 & 7 & 5 &   11  & 2\\
    \hline
    \label{table:num}
  \end{tabular}

\end{table}

\begin{figure*}[!ht]
  \centering
    \includegraphics[scale = 1]{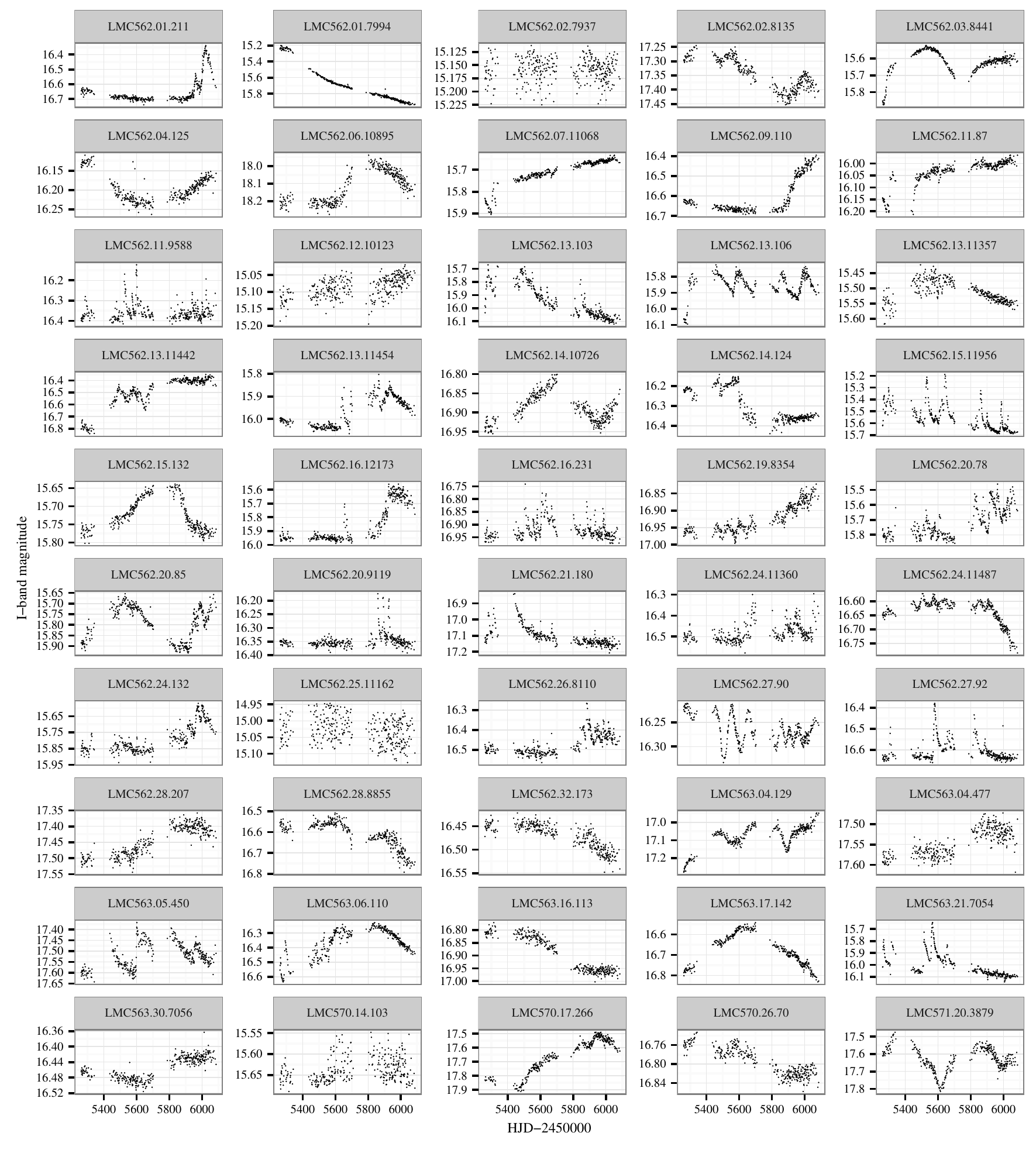}
    \caption{Time series of the selected 50 BeSC in the OGLE-IV 
    Gaia south ecliptic pole field.
   Observations were sampled in a window close 900 days.}
  \label{figure:selectedBeLightCurves}
\end{figure*}

It is seen that the majority of the BeSC selected are Type-1 stars.
This reflects the useful effect of considering the LMC BeSC subsample
as part of the training sample: in our Galaxy the amount of
outbursting stars is much smaller than in the LMC, but since the GSEP
field is near the LMC centre ($\sim 5^\circ$), it is
  expected to find outbursting BeSC. It is also worth noting that the
presence in our catalogue of objects showing a brightness
discontinuity of magnitude (Type-2 stars). Again, since these objects
are observed in the direction of the GSEP, it is more probable that
they are members of the LMC than of the Galaxy, where this type of
variability for BeSC has never been detected. Spectroscopic follow-up
of these stars are needed to confirm their \textit{Be} nature.  Figure
\ref{figure:selectedBeLightCurves} shows the time series of the 50
stars selected using the random forests algorithm and the colour
criteria.

A fraction of the stars discarded from those initially selected as
BeSC by our random forest procedure are periodic stars, as reported by
the OGC. This fact gives more evidence that they are actually
SPV, LPV, or non-periodic variable stars.  LMC562.32.265 and
LMC562.23.11510 are between the stars
discarded by the colour criterion. The light curves of these non-periodic variables
had been shown in \citet[Fig. 7]{soszynski_optical_2012}. 
They are very similar to these of
Type-1 and Type-2 BeSC, but their (V-I) colours are redder than the
expected for \textit{Be} stars.

\section{Infrared colours of the BeSC} \label{sect:Infrared colours of the BeSC}
\begin{figure}[!ht]
  \centering
    \includegraphics[width=\hsize]{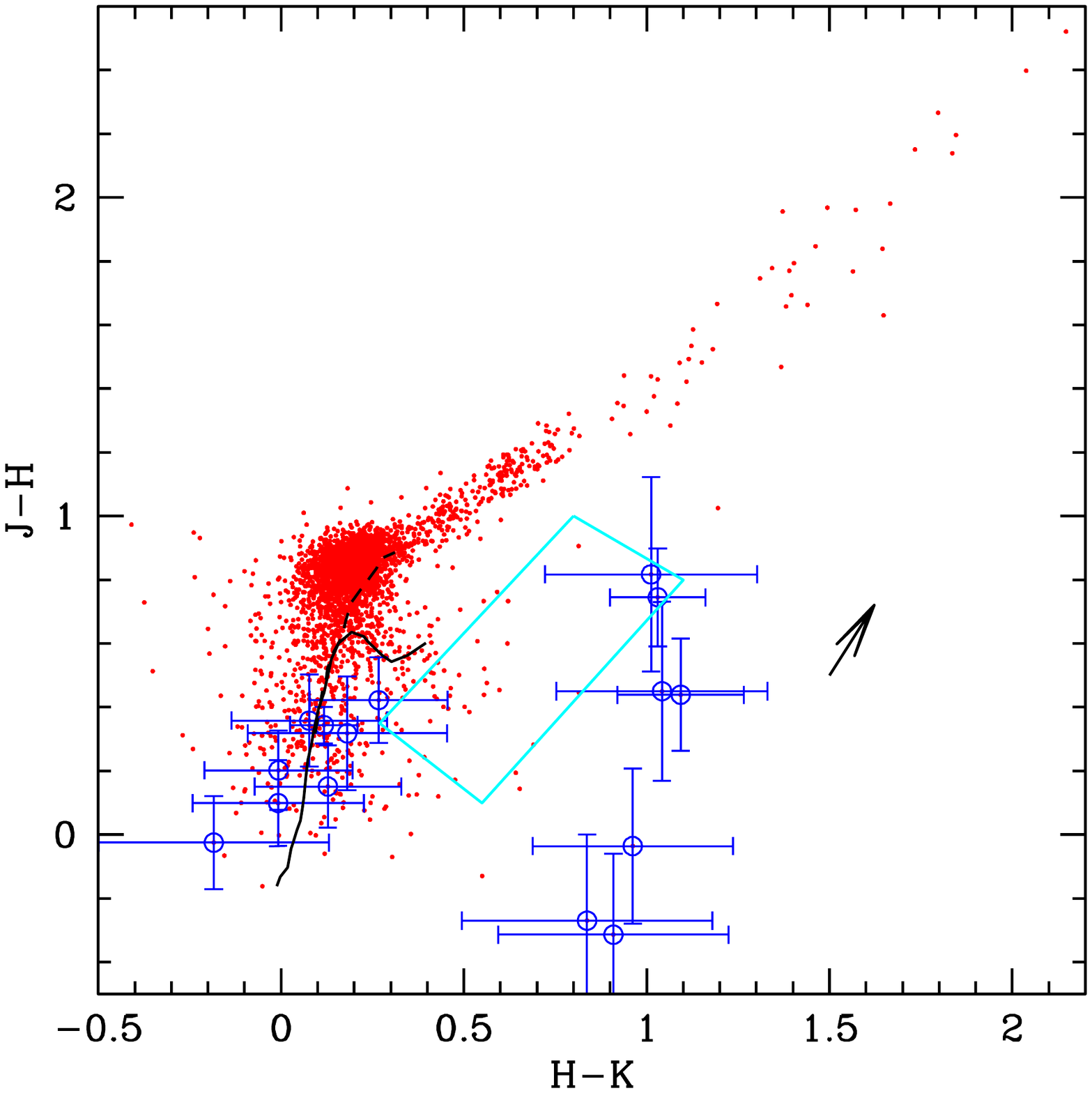}
    \caption{2MASS colour-colour diagram for the variable stars (red dots) reported 
    by \citet[]{soszynski_optical_2012} and the 15 BeSC with 2MASS 
    counterpart (blue open circles). The standard sequences from \citet{Bessell} 
    are shown in solid line (main sequence) and in dashed line (giant sequence). 
    The loci of HAeBe stars is represented by the cyan box. Classical \textit{Be} stars 
    are located in the region near the blue end of the main sequence 
    \citep{Hernandez_2005}. The arrow represents the reddening vector for 
    2 magnitudes of visual extinction.}
  \label{figure:colorBeSC}
\end{figure}

Using 2MASS and WISE catalogues we explore the infrared properties of 
our selected sample of BeSC.  
Most of the 50 BeSC do not have reliable photometry in the 2MASS catalogue. 
Figure \ref{figure:colorBeSC} shows the distribution of 15 BeSC in the 2MASS 
colour-colour diagram. About 8 BeSC have 2MASS colours with different 
levels of reddening. There are 4 stars (LMC563.04.129, LMC563.17.142, LMC570.26.70,  and
LMC570.17.266) that fall in the HAeBe region defined by 
\citet{Hernandez_2005} and have WISE colours consistent with HAeBe stars 
\citep{Koenig,Hernandez_p}. These stars are HAeBe candidates that could be 
surrounded by an optically thick accretion disk. The detection of HAeBe 
stars in the LMC has been reported previously \citep[e.g. ][]{Hatano_2006}. 
Finally, there are 3 stars that fall below the HAeBe region (LMC562.13.11454, 
LMC562.26.8110, and LMC562.13.11454); these stars can be high mass objects 
(O type or early B) surrounded by a cool circumstellar  envelope that produces 
excess at K band. Spectroscopic observations are necessary to  reveal the 
nature of these objects.  Despite the small sample of BeSC with infrared colours, 
apparently there is no relation between the morphological type of BeSC in table 
\ref{table:catalog} and the location on the 2MASS colour-colour diagram. 

\section{Conclusions \label{sect:conclusions}}

In this work we presented and tested a new set of robust
  features for the supervised classification of variable stars and
  presented a new catalogue of 50 \textit{Be} star candidates, four of which had infrared
  colours that were consistent with Herbig Ae/Be stars. 

We presented a new set of features and showed their
  usefulness for the automatic classification of variable
stars. This features are statistical parameters 
computed based on the I band magnitude density of
the light curves that are robust to the presence
  of outliers. These parameters quantify the
location, scale, skewness, tail weight, and smoothness of the
  magnitude density.
  
In order to prove the usefulness of our proposed set of
  features, we trained state-of-the-art classifiers on a sample
  of light curves from diverse variability types: Cepheids, $\delta$ 
  Scuti, eclipsing binaries, long period variables, type II Cepheids, 
  RR Lyr\ae,{} and \textit{Be} star candidates. 
  We tuned and tested the performance of classification
  trees and random forests along with K-nearest neighbours, support vector
  machines and gradient boosted trees via a grid search, 
  10-fold cross-validation, and the mean $F_1$ score based on 
  normalised confusion matrices as performance
  metric. Our classifiers yielded correct classifications
with high probability, which shows that our proposed set of features
can be used to characterise different variability types. We
found that the random forest classifier produces the best
results.

We used the trained random forest classifier to look for \textit{Be} star candidates in a subset of 1473 variable stars classified as Other in the  OGLE-IV Gaia south ecliptic pole field field catalogue. After further selection using colour criteria, we present a new catalogue of 50 \textit{Be} star candidates. Despite the necessity of
a spectroscopic follow-up to confirm the presence of Balmer emission lines, and consequently the \textit{Be} nature of these stars,  
their optical and infrared colours correspond to
the expected for \textit{Be} stars, except for four stars that have colours consistent with those
of Herbig Ae/Be variables. Because there are BeSC in our selected sample 
showing in their light curves jumps or brightness discontinuities
never observed in the Milky Way (Type 2 stars), this suggests that probably they 
 belong to the Large Magellanic Cloud.

\begin{acknowledgements}
AGV and BES acknowledge financial support from Vicerrector\'ia de investigaciones, Universidad
de los Andes, through programme: Asignaci\'on de recursos destinados a la finalizaci\'on de proyectos conducentes a la obtenci\'on de nuevo conocimiento.
\end{acknowledgements}

\begin{appendix}
\section{Uniform prior probability, sensitivity, and specificity estimation \label{appendix:UniformPrior}}
Here we show why the normalisation of the confusion matrix that we
perform is equivalent to assigning a uniform \textit{a priori}
probability distribution to the observations of members of each
variability class. First, we need to fix some notation.  A
classifier $g$ is a function that assigns to each vector $\vec{x}$ a
class $i\in\{1,\dots,M\}$. The value $P(\vec{x}, i)$ is the probability that a
feature vector $\vec{x}$ that corresponds to the class $i$ is
observed. The recall seeks to estimate $P(g(\vec{x})=i|i)$, and the
precision $P(i|g(\vec{x})=i)$ when the sample is representative of
the object population. Since the sample considered in this work is
surely not representative of the star populations, we need to assign
subjectively a priori probabilities to the different
variability classes. Because to our best knowledge there are no
studies in this regard, we choose a uniform prior, that is, $P(i) =
1/7$ for all classes.  We can write in the case of uniform a
  priori probabilities
\begin{equation}
  P(i|g(\vec{x})=i) = \frac{P(g(x)=i|i)}{\sum_{j}P(g(x)=i|j)}.
\end{equation}
For each $i$ and $j$ in the k-th iteration of the 10-fold cross-validation, we estimate $P(g(\vec{x})=i|j)$ with
\begin{equation}
\label{eq:recallEst}
  \hat{P}^{(k)}(g(\vec{x})=i|j) = \frac{C_{ij}^{(k)}}{\sum_{k}C_{kj}^{(k)}} 
,\end{equation}
where $C^{(k)}$ is the confusion matrix of the k-th holdout
sample. The value $\hat{P}^{(k)}(g(\vec{x})=i|j)$ is the $ij$ entry of the
normalised confusion matrix, which we call $\hat{C}^{(k)}$. For the
i-th class, the estimated recall for the k-th iteration, as given by
\ref{eq:recallEst}, is just $\hat{C}^{(k)}_{ii}$. Our estimator of the
precision in each iteration is
\begin{align}
  \hat{P}^{(k)}(i|g(\vec{x})=i)&= \frac{\hat{P}^{(k)}(g(x)=i|i)}{\sum_{j}\hat{P}^{(k)}(g(x)=i|j)},\\
  &= \frac{\hat{C}^{(k)}_{ii}}{\sum_{j} \hat{C}^{(k)}_{ij}}
\end{align}
which is the precision calculated with the normalised confusion
matrix. Finally, our cross-validation estimators of
$P(i|g(\vec{x})=i)$, and $P(g(\vec{x})=i|i)$ are just the average of
the estimates over the folds, that is,
\begin{align}
  \hat{P}(i|g(\vec{x})=i) &= \frac{1}{10} \sum_{k} \hat{P}^{(k)}(i|g(\vec{x})=i)\\
  P(g(\vec{x})=i|i) &= \frac{1}{10} \sum_{k} \hat{P}^{(k)}(g(\vec{x})= i|i).
\end{align}

\section{Other classifiers \label{appendix:otherclassifiers}}
\subsubsection{K-nearest neighbours (KNN)}

The KNN classifier was first proposed by \citet{fix_discriminatory_1951} and 
republished by \citet{silverman_e._1989}. This algorithm is based on
the observation that the examples of one class are close to each other
and that it is possible to classify one example based on its nearest
neighbours. Given a fixed integer, $k$, this rule assigns to each point
in feature space the class to which the majority of its $k$ nearest
neighbours belongs. It is possible to show that KNN converges to the
best possible classification rule for a given set of features as the
number of examples $N\to\infty$ as long as $k/N\to 0$. Despite its
simplicity, KNN has been shown to be a competitive rule in the sense that
it achieves accuracies comparable to those of more sophisticated
decision rules, and only one parameter, the number $k$ of neighbours,
needs to be tuned.

There exist weighted and bagged schemes of KNN. In weighting schemes,
to each of the k nearest neighbours is given a different weight in the
final decision. Bagging (short for bootstrap aggregating) consists of
averaging the decision of several KNN classifiers trained with
bootstrap samples of the original training sample, i.e. samples of the
same size taken randomly with replacement from the original training
sample. It has been shown that this reduces over-fitting and variance
\citep[]{breiman_bagging_1996}. \citet{samworth_optimal_2012} showed
that bagging is asymptotically equivalent to a weighted scheme and
that there exists an optimal weighting scheme. We compare unweighted,
optimal weighted (as shown by \citet{samworth_optimal_2012}), and
bagged KNN classifiers with the FNN package
\citep[][]{beygelzimer_fnn:_2013}, which provides a fast
implementation for these methods.\\
We scale the data so that each feature has standard deviation 1 and mean 0 and
assess the performance of the model for 5 values:
k = 1; 3; 5; 7; 9, finding that the best performance is achieved for
low values of k and choose k = 1.

\subsubsection{Support vector machines (SVM)}

The SVM were first proposed by \citet{cortes_support-vector_1995} and a
complete introduction to the topic can be found in
\citet{cristianini_introduction_2000}. The SVM are binary classifiers that
divide a transformed version of the feature space into two
regions by finding the hyper-plane that separates data of both classes
with maximal margin. Data are transformed hoping that in the high-dimensional space they are linearly separable. The maximal margin
hyper-plane can be found by solving a convex optimisation problem for
which efficient solvers are available and it includes a
misclassification cost term that is controlled by a single parameter
$C$. The transformation of the data into the high-dimensional space
does not have to be known because the convex optimisation problem can
only be solved by using the matrix of dot products in the high-dimensional space, which can be calculated directly using kernel
functions. Consequently, the choice of kernel function is crucial for
the performance of SVM. One of the most popular kernel functions is
the radial basis kernel,
\begin{equation}
  \label{eq:radialBasis}
  K(x,y) = e^{-\gamma\lVert x-y\rVert^2},
\end{equation}
which has only one free parameter, $\gamma$. We tune the cost
parameter $C$ and $\gamma$.  In order to perform a $M$ class
classification with SVM there are two popular approaches. The first
one is called one-against-one and it consists of training
$\frac{M(M-1)}{2}$ SVM that distinguish between each pair of
classes. The final decision is to choose the class selected
  most often by the classifiers. The second one is called
one-against-all and $M$ SVM are trained to distinguish between each
class and the data non-belonging to that class. The decision is to
select the class chosen by its classifier with the largest
margin. One-against-one has proven to be faster and both approaches
yield similar classification performances
\citep[][]{hsu_comparison_2002}.

We use the interface to the \texttt{libsvm} implementation of SVM
\citep{chang_libsvm:_2011} of the \texttt{e1071} package
\citep[][]{meyer_e1071:_2015} and the wrapper function from the
package \texttt{caret}. \\

Before adjusting the SVM, data are scaled so that each 
feature has a standard deviation of 1 and a mean of 0. The 
parameters $\gamma$ and C are selected by cross-validation as 0.04 
and $2^{11}$, respectively. Candidates considered for $\gamma$ were 
equally spaced numbers between the reciprocals of the 0.1 and 0.9 
percentiles of the interpoint distance distribution in the scaled 
feature space, while candidate values for C were powers of 2.

\subsubsection{Gradient-boosted trees \label{sect:MethBoostedTrees}}

Gradient boosting was proposed by
\citet{friedman_greedy_2001}. In a similar fashion to random
  forests, it is based on the idea that a set of weak classifiers
  (classification trees) can be chosen to conform a strong
  classifier. In this case, each classification tree is built in a
  stagewise greedy manner, that is, each tree is built sequentially to
  maximise the decrease of a loss function associated with
  misclassification. During the training process, each tree is  assigned different weight in the final decision of the classifier, whose
  final decision is the result of the weighted voting among the
  classification trees.

  We use the implementation of the \texttt{xgboost} package
  \citep[][]{chen_xgboost:_2015} and several parameters need to be
  tuned. The learning rate, the number of trees, and their depth can be
  modified. The number of trees that are built is modified by the
  parameter \texttt{nrounds}. The learning rate modifies the
  contribution that each tree makes to the classifier and can be
  modified by changing between 0 and 1 the parameter \texttt{eta}. 
  A smaller value \texttt{eta} makes the training more conservative, which means that a larger number of \texttt{nrouds} is needed. The
  depth of each tree is controlled by the parameter
  \texttt{max\_depth}. We tune both \texttt{nrouds} and
  \texttt{max\_depth} and left \texttt{eta}
  fixed to its default value of 0.3.\\\\
  By grid search, the number of trees that are grown was set
    to nround = 100, while the maximum depth of the trees was chosen
    as \texttt{max\_depth=7}.

\end{appendix} 

\bibliographystyle{aa}
\bibliography{classificationArticle,rPackages,classificationArticleVersionProfe,revisionAdolfo}

\begin{thebibliography}{71}
\expandafter\ifx\csname natexlab\endcsname\relax\def\natexlab#1{#1}\fi

\bibitem[{Bass(2016)}]{Bass}
Bass, G.and~Borne, K. 2016, MNRAS, 459, 3721

\bibitem[{Bessel \& Brett(1988)}]{Bessell}
Bessel, M.~S. \& Brett, J.~M. 1988, PASP, 100, 1134

\bibitem[{Beygelzimer {et~al.}(2013)Beygelzimer, Kakadet, Langford, Arya,
  Mount, \& Li}]{beygelzimer_fnn:_2013}
Beygelzimer, A., Kakadet, S., Langford, J., {et~al.} 2013, FNN: Fast nearest
  neighbor search algorithms and applications., r package version 1.1

\bibitem[{Biau {et~al.}(2008)Biau, Devroye, \& Lugosi}]{biau_consistency_2008}
Biau, G., Devroye, L., \& Lugosi, G. 2008, The Journal of Machine Learning
  Research, 9, 2015

\bibitem[{Breiman(1996)}]{breiman_bagging_1996}
Breiman, L. 1996, Machine Learning, 24, 123

\bibitem[{Breiman(2001)}]{breiman_random_2001}
Breiman, L. 2001, Machine Learning, 45, 5

\bibitem[{Breiman {et~al.}(1984)Breiman, Friedman, Stone, \&
  Olshen}]{breiman_classification_1984}
Breiman, L., Friedman, J., Stone, C.~J., \& Olshen, R.~A. 1984, Classification
  and {Regression} {Trees}, 1st edn. (New York, N.Y.: Chapman and Hall/CRC)

\bibitem[{Brys {et~al.}(2004)Brys, Hubert, \& Struyf}]{brys_robust_2004}
Brys, G., Hubert, M., \& Struyf, A. 2004, Journal of Computational and
  Graphical Statistics, 13

\bibitem[{Brys {et~al.}(2006)Brys, Hubert, \& Struyf}]{brys_robust_2006}
Brys, G., Hubert, M., \& Struyf, A. 2006, Computational statistics \& data
  analysis, 50, 733

\bibitem[{Chang \& Lin(2011)}]{chang_libsvm:_2011}
Chang, C.-C. \& Lin, C.-J. 2011, ACM Transactions on Intelligent Systems and
  Technology (TIST), 2, 27

\bibitem[{Chen {et~al.}(2015)Chen, He, \& Benesty}]{chen_xgboost:_2015}
Chen, T., He, T., \& Benesty, M. 2015, xgboost: {Extreme} {Gradient}
  {Boosting}, r package version 0.4-2

\bibitem[{Collins(1987)}]{CO}
Collins, G.~W. 1987, Physics of Be stars, ed. A.~Slettebak \& T.~P. Snow, Proc.
  IAU Coll. 92 (Cambridge University Press)

\bibitem[{Cortes \& Vapnik(1995)}]{cortes_support-vector_1995}
Cortes, C. \& Vapnik, V. 1995, Machine Learning, 20, 273

\bibitem[{Cristianini \& Shawe-Taylor(2000)}]{cristianini_introduction_2000}
Cristianini, N. \& Shawe-Taylor, J. 2000, An introduction to support vector
  machines and other kernel-based learning methods (Cambridge University press)

\bibitem[{Deb \& Singh(2009)}]{deb_light_2009}
Deb, S. \& Singh, H.~P. 2009, Astronomy and Astrophysics, 507, 1729

\bibitem[{Debosscher {et~al.}(2007)Debosscher, Sarro, Aerts, Cuypers,
  Vandenbussche, Garrido, \& Solano}]{debosscher_automated_2007}
Debosscher, J., Sarro, L.~M., Aerts, C., {et~al.} 2007, A\&A, 475, 1159

\bibitem[{Fix \& Hodges~Jr(1951)}]{fix_discriminatory_1951}
Fix, E. \& Hodges~Jr, J.~L. 1951, Discriminatory analysis-nonparametric
  discrimination: consistency properties, Tech. rep., DTIC Document

\bibitem[{Friedman(2001)}]{friedman_greedy_2001}
Friedman, J.~H. 2001, The Annals of Statistics, 29, 1189

\bibitem[{Graczyk {et~al.}(2011)Graczyk, Soszy\'nski, Poleski, Pietrzyński,
  Udalski, Szymański, Kubiak, Wyrzykowski, \& Ulaczyk}]{graczyk_optical_2011}
Graczyk, D., Soszy\'nski, I., Poleski, R., {et~al.} 2011, Acta Astron., 61, 103

\bibitem[{Hampel {et~al.}(1986)Hampel, Ronchetti, Rousseeuw, \&
  Stahel}]{hampel_robust_1986}
Hampel, F.~R., Ronchetti, E.~M., Rousseeuw, P.~J., \& Stahel, W.~A. 1986,
  Robust statistics: the approach based on influence functions (John Wiley \&
  Sons)

\bibitem[{Hastie {et~al.}(2009)Hastie, Tibshirani, \&
  Friedman}]{hastie_elements_2009}
Hastie, T., Tibshirani, R., \& Friedman, J. 2009, The {Elements} of
  {Statistical} {Learning}, Springer {Series} in {Statistics} (New York, NY:
  Springer New York)

\bibitem[{Hatano {et~al.}(2006)Hatano, Kadowaki, Nakajima, Tamura, Nagata,
  Sugitani, \& Tanab\'e}]{Hatano_2006}
Hatano, H., Kadowaki, R., Nakajima, Y., {et~al.} 2006, AJ, 132, 2653

\bibitem[{Hern\'andez {et~al.}(2005)Hern\'andez, Calvet, Hartmann, Briceño,
  Sicilia-Aguilar, \& Berlind}]{Hernandez_2005}
Hern\'andez, J., Calvet, N., Hartmann, L., {et~al.} 2005, AJ, 129, 856

\bibitem[{Hern\'andez {et~al.}(2017)Hern\'andez, Villareal, Calvet, Mauc\'o,
  Ballesteros, Galv\'an, \& Olgu\'in}]{Hernandez_p}
Hern\'andez, J., Villareal, L., Calvet, N., {et~al.} 2017, In prep.

\bibitem[{Hsu \& Lin(2002)}]{hsu_comparison_2002}
Hsu, C.-W. \& Lin, C.-J. 2002, Neural Networks, IEEE Transactions on, 13, 415

\bibitem[{Huber(1964)}]{huber_robust_1964}
Huber, P.~J. 1964, The Annals of Mathematical Statistics, 35, 73

\bibitem[{Huber \& Ronchetti(2009)}]{huber_robust_2009}
Huber, P.~J. \& Ronchetti, E.~M. 2009, Robust {Statistics}, 2nd edn. (Hoboken,
  N.J: Wiley)

\bibitem[{Hubert \& Floquet(1998)}]{HU8}
Hubert, A.~M. \& Floquet, M. 1998, A\&A, 335, 565

\bibitem[{Hubert {et~al.}(2000)Hubert, Floquet, \& Zorec}]{HU}
Hubert, A.~M., Floquet, M., \& Zorec, J. 2000, Proc. IAU Coll. 175, The Be
  Phenomenon in Early-Type Stars., ed. M.~A. Smith, H.~F. Henrichs, \&
  J.~Fabregat, Proc. IAU Coll. 92 (Astron. Soc. Pac.)

\bibitem[{Khun(2016)}]{wing_caret:_2016}
Khun, M. 2016, caret: Classification and Regression Training, , r package
  version 6.0-6.4

\bibitem[{Kim {et~al.}(2014)Kim, Protopapas, Bailer-Jones, Byun, Chang,
  Marquette, \& Shin}]{kim_epoch_2014}
Kim, D.-W., Protopapas, P., Bailer-Jones, C. A.~L., {et~al.} 2014, A\&A, 566,
  A43

\bibitem[{Koenig(2014)}]{Koenig}
Koenig, X. P. \&~Leisawitz, D.~T. 2014, ApJ, 791, 131

\bibitem[{Krijthe(2015)}]{rtsne_package}
Krijthe, J. 2015, Rtsne: T-Distributed Stochastic Neighbor Embedding using
  Barnes-Hut Implementation, r package version 0.10

\bibitem[{Liaw \& Wiener(2002)}]{liaw_classification_2002}
Liaw, A. \& Wiener, M. 2002, R News, 2, 18

\bibitem[{Mennickent {et~al.}(2002)Mennickent, Pietrzy\'nski, Gieren, \&
  Szewczyk}]{MEN2}
Mennickent, R.~E., Pietrzy\'nski, G., Gieren, W., \& Szewczyk, O. 2002, A\&A,
  393, 887

\bibitem[{Meyer {et~al.}(2015)Meyer, Dimitriadou, Hornik, Weingessel, \&
  Leisch}]{meyer_e1071:_2015}
Meyer, D., Dimitriadou, E., Hornik, K., Weingessel, A., \& Leisch, F. 2015,
  e1071: {Misc} {Functions} of the {Department} of {Statistics}, {Probability}
  {Theory} {Group} ({Formerly}: {E}1071), {TU} {Wien}, r package version 1.6-7

\bibitem[{Mowlavi(2014)}]{mowlavi_searching_2014}
Mowlavi, N. 2014, A\&A, 568, 78

\bibitem[{Park {et~al.}(2013)Park, Oh, \& Kim}]{park_classification_2013}
Park, M., Oh, H.-S., \& Kim, D. 2013, PASP, 125, 470

\bibitem[{Pawlak {et~al.}(2013)Pawlak, Graczyk, Soszy\'nski, Pietrukowicz,
  Poleski, Udalski, Szymański, Kubiak, Pietrzyński, Wyrzykowski, Ulaczyk,
  Kozłowski, \& Skowron}]{pawlak_eclipsing_2013-1}
Pawlak, M., Graczyk, D., Soszy\'nski, I., {et~al.} 2013, Acta Astron., 63, 323

\bibitem[{Pichara {et~al.}(2016)Pichara, Protopapas, \& Le\'on}]{Pichara}
Pichara, K., Protopapas, P., \& Le\'on, D. 2016, ApJ, 819, 18

\bibitem[{Poleski {et~al.}(2010)Poleski, Soszy\'nski, Udalski, Szymański,
  Kubiak, Pietrzyński, Wyrzykowski, Szewczyk, \&
  Ulaczyk}]{poleski_optical_2010}
Poleski, R., Soszy\'nski, I., Udalski, A., {et~al.} 2010, Acta Astron., 60, 1

\bibitem[{R~Core~Team(2015)}]{rcore2015}
R~Core~Team, . 2015, R: A Language and Environment for Statistical Computing
  (Vienna, Austria: R Foundation for Statistical Computing)

\bibitem[{Rivinius {et~al.}(2013)Rivinius, Carciofi, \& Martayan}]{Ri}
Rivinius, T., Carciofi, A.~C., \& Martayan, C. 2013, A\&ARv, 21, 69

\bibitem[{Sabogal {et~al.}(2014)Sabogal, García-Varela, \&
  Mennickent}]{sabogal_search_2014}
Sabogal, B.~E., García-Varela, A., \& Mennickent, R.~E. 2014, PASP, 126, 219

\bibitem[{Sabogal {et~al.}(2005)Sabogal, Mennickent, Pietrzy\'nski, \&
  Gieren}]{Sabogal2005}
Sabogal, B.~E., Mennickent, R.~E., Pietrzy\'nski, G., \& Gieren, W. 2005,
  MNRAS, 361, 1055

\bibitem[{Sabogal {et~al.}(2008)Sabogal, Mennickent, Pietrzyński, García,
  Gieren, \& Kolaczkowski}]{sabogal_catalogue_2008}
Sabogal, B.~E., Mennickent, R.~E., Pietrzyński, G., {et~al.} 2008, A\&A, 478,
  659

\bibitem[{Samworth(2012)}]{samworth_optimal_2012}
Samworth, R.~J. 2012, The Annals of Statistics, 40, 2733

\bibitem[{Sarro {et~al.}(2009)Sarro, Debosscher, López, \&
  Aerts}]{sarro_automated_2009}
Sarro, L.~M., Debosscher, J., López, M., \& Aerts, C. 2009, A\&A, 494, 739

\bibitem[{Schlafly \& Finkbeiner(2011)}]{Schlafly}
Schlafly, E. \& Finkbeiner, D. 2011, ApJ, 737, 103

\bibitem[{Silverman \& Jones(1989)}]{silverman_e._1989}
Silverman, B.~W. \& Jones, M.~C. 1989, International Statistical Review/Revue
  Internationale de Statistique, 57, 233

\bibitem[{Soszy\'nski {et~al.}(2011{\natexlab{a}})Soszy\'nski, Dziembowski,
  Udalski, Poleski, Szymański, Kubiak, Pietrzyński, Wyrzykowski, Ulaczyk,
  Kozłowski, \& Pietrukowicz}]{soszynski_optical_2011-2}
Soszy\'nski, I., Dziembowski, W.~A., Udalski, A., {et~al.} 2011{\natexlab{a}},
  Acta Astron., 61, 1

\bibitem[{Soszy\'nski {et~al.}(2008{\natexlab{a}})Soszy\'nski, Poleski,
  Udalski, Szymański, Kubiak, Pietrzyński, Wyrzykowski, Szewczyk, \&
  Ulaczyk}]{soszynski_optical_2008-1}
Soszy\'nski, I., Poleski, R., Udalski, A., {et~al.} 2008{\natexlab{a}}, Acta
  Astron., 58, 163

\bibitem[{Soszy\'nski {et~al.}(2010{\natexlab{a}})Soszy\'nski, Poleski,
  Udalski, Szymański, Kubiak, Pietrzyński, Wyrzykowski, Szewczyk, \&
  Ulaczyk}]{soszynski_optical_2010-2}
Soszy\'nski, I., Poleski, R., Udalski, A., {et~al.} 2010{\natexlab{a}}, Acta
  Astron., 60, 17

\bibitem[{Soszy\'nski {et~al.}(2011{\natexlab{b}})Soszy\'nski, Udalski,
  Pietrukowicz, Szymański, Kubiak, Pietrzyński, Wyrzykowski, Ulaczyk,
  Poleski, \& Kozłowski}]{soszynski_optical_2011}
Soszy\'nski, I., Udalski, A., Pietrukowicz, P., {et~al.} 2011{\natexlab{b}},
  Acta Astron., 61, 285

\bibitem[{Soszy\'nski {et~al.}(2013{\natexlab{a}})Soszy\'nski, Udalski,
  Pietrukowicz, Szymański, Kubiak, Pietrzyński, Wyrzykowski, Ulaczyk,
  Poleski, \& Kozłowski}]{soszynski_optical_2013}
Soszy\'nski, I., Udalski, A., Pietrukowicz, P., {et~al.} 2013{\natexlab{a}},
  Acta Astron., 63, 37

\bibitem[{Soszy\'nski {et~al.}(2012)Soszy\'nski, Udalski, Poleski, Kozlowski,
  Wyrzykowski, Pietrukowicz, Szyma\'nski, Kubiak, Pietrzynski, Ulaczyk, \&
  {others}}]{soszynski_optical_2012}
Soszy\'nski, I., Udalski, A., Poleski, R., {et~al.} 2012, Acta Astron., 62, 219

\bibitem[{Soszy\'nski {et~al.}(2010{\natexlab{b}})Soszy\'nski, Udalski,
  Szymański, Kubiak, Pietrzyński, Wyrzykowski, Ulaczyk, \&
  Poleski}]{soszynski_optical_2010}
Soszy\'nski, I., Udalski, A., Szymański, M.~K., {et~al.} 2010{\natexlab{b}},
  Acta Astron., 60, 165

\bibitem[{Soszy\'nski {et~al.}(2008{\natexlab{b}})Soszy\'nski, Udalski,
  Szymański, Kubiak, Pietrzyński, Wyrzykowski, Szewczyk, Ulaczyk, \&
  Poleski}]{soszynski_optical_2008}
Soszy\'nski, I., Udalski, A., Szymański, M.~K., {et~al.} 2008{\natexlab{b}},
  Acta Astron., 58, 293

\bibitem[{Soszy\'nski {et~al.}(2009{\natexlab{a}})Soszy\'nski, Udalski,
  Szymański, Kubiak, Pietrzyński, Wyrzykowski, Szewczyk, Ulaczyk, \&
  Poleski}]{soszynski_optical_2009-1}
Soszy\'nski, I., Udalski, A., Szymański, M.~K., {et~al.} 2009{\natexlab{a}},
  Acta Astron., 59, 1

\bibitem[{Soszy\'nski {et~al.}(2009{\natexlab{b}})Soszy\'nski, Udalski,
  Szymański, Kubiak, Pietrzyński, Wyrzykowski, Szewczyk, Ulaczyk, \&
  Poleski}]{soszynski_optical_2009}
Soszy\'nski, I., Udalski, A., Szymański, M.~K., {et~al.} 2009{\natexlab{b}},
  Acta Astron., 59, 239

\bibitem[{Soszy\'nski {et~al.}(2010{\natexlab{c}})Soszy\'nski, Udalski,
  Szymański, Kubiak, Pietrzyński, Wyrzykowski, Ulaczyk, \&
  Poleski}]{soszynski_optical_2010-1}
Soszy\'nski, I., Udalski, A., Szymański, M.~K., {et~al.} 2010{\natexlab{c}},
  Acta Astron., 60, 91

\bibitem[{Soszy\'nski {et~al.}(2011{\natexlab{c}})Soszy\'nski, Udalski,
  Szymański, Kubiak, Pietrzyński, Wyrzykowski, Ulaczyk, Poleski, Kozłowski,
  \& Pietrukowicz}]{soszynski_optical_2011-1}
Soszy\'nski, I., Udalski, A., Szymański, M.~K., {et~al.} 2011{\natexlab{c}},
  Acta Astron., 61, 217

\bibitem[{Soszy\'nski {et~al.}(2013{\natexlab{b}})Soszy\'nski, Udalski,
  Szymański, Kubiak, Pietrzyński, Wyrzykowski, Ulaczyk, Poleski, Kozłowski,
  Pietrukowicz, \& Skowron}]{soszynski_optical_2013-1}
Soszy\'nski, I., Udalski, A., Szymański, M.~K., {et~al.} 2013{\natexlab{b}},
  Acta Astron., 63, 21

\bibitem[{Staudte \& Sheather(1990)}]{staudte_robust_1990}
Staudte, R.~G. \& Sheather, S.~J. 1990, Robust estimation and testing, Wiley
  {Series} in {Probability} and {Statistics} (John Wiley \& Sons)

\bibitem[{Therneau {et~al.}(2015)Therneau, Atkinson, \&
  Ripley}]{therneau_rpart:_2015}
Therneau, T., Atkinson, B., \& Ripley, B. 2015, rpart: Recursive Partitioning
  and Regression Trees, r package version 4.1-10

\bibitem[{Udalski(2004)}]{udalski_optical_2004}
Udalski, A. 2004, Acta Astron., 53, 291

\bibitem[{Udalski {et~al.}(2015)Udalski, Szyma\'nski, \&
  Szyma\'nski}]{Udalski2015}
Udalski, A., Szyma\'nski, M.~K., \& Szyma\'nski, G. 2015, Acta Astron., 65, 138

\bibitem[{Van~der Maaten \& Hinton(2008)}]{van_der_maaten_visualizing_2008}
Van~der Maaten, L. \& Hinton, G. 2008, The Journal of Machine Learning
  Research, 9, 85

\bibitem[{Venables \& Ripley(2013)}]{venables_modern_2013}
Venables, W.~N. \& Ripley, B.~D. 2013, Modern applied statistics with
  {S}-{PLUS} (Springer Science \& Business Media)

\bibitem[{Von~Neumann(1941)}]{von_neumann_distribution_1941}
Von~Neumann, J. 1941, The Annals of Mathematical Statistics, 12, 367

\bibitem[{Wisniewski \& Bjorkman(2006)}]{Wisniewski}
Wisniewski, J.~P. \& Bjorkman, K.~S. 2006, ApJ, 652, 458

\end{thebibliography}
\end{document}